\documentclass[12pt,a4paper]{article}
\pdfoutput=1

\usepackage{geometry}
\usepackage[numbers,sort&compress]{natbib}
\usepackage{amsmath}
\usepackage{amssymb}
\usepackage[dvips]{graphicx}
\usepackage{pstricks}
\usepackage{bm}
\usepackage{pbox}
\usepackage{placeins}
\usepackage{graphicx}
\usepackage{caption}
\usepackage{subcaption}
\usepackage[T1]{fontenc}
\usepackage{footnote}
\usepackage{pdfpages}
\usepackage{hhline}
\usepackage{multirow}
\usepackage{enumitem}
\usepackage[nottoc,notlot,notlof]{tocbibind}
\usepackage[title,titletoc]{appendix}
\usepackage{dsfont}
\allowdisplaybreaks

\usepackage{cases}

\newcommand{\DR}{\textcolor{red}}

\definecolor{MyDarkBlue}{rgb}{0.1, 0.1, 0.8} 
\definecolor{MyLightBlue}{rgb}{0.22,0.51,0.9}
\definecolor{MyGreen}{rgb}{0.0, 0.5, 0.0}
\definecolor{BrickRed}{rgb}{0.8, 0.25, 0.33}
\usepackage[colorlinks=true,linkcolor=blue,citecolor=MyDarkBlue,
urlcolor=MyLightBlue,bookmarksnumbered=true,bookmarksopen]{hyperref}
\hypersetup{colorlinks, citecolor=blue,linkcolor=red, urlcolor=MyLightBlue}

\begin{document}
\vspace*{-0.2in}
\begin{flushright}
\end{flushright}

\begin{center}
 {\Large \bf Monopoles, Exotic states and Muon $g-2$ in TeV scale
Trinification}  
\end{center}

\vspace{0.5cm}
\renewcommand{\thefootnote}{\fnsymbol{footnote}}
\begin{center}
{\large
{}~\textbf{Digesh Raut}$^{a,b}$\footnote{ E-mail: \textcolor{MyLightBlue}{draut@udel.edu}},
{}~\textbf{Qaisar Shafi}$^{b}$\footnote{ E-mail: \textcolor{MyLightBlue}{qshafi@udel.edu}}
 and 
{}~\textbf{Anil Thapa}$^{c}$\footnote{ E-mail: \textcolor{MyLightBlue}{wtd8kz@virginia.edu}} 
}
\vspace{0.5cm}

$^{a}${\em Physics Department,  Washington College,  Chestertown,  MD 21620, USA }

$^{b}${\em Bartol Research Institute, Department of Physics and Astronomy,University of Delaware, Newark DE 19716, USA }

$^{c}${\em Department of Physics, University of Virginia, Charlottesville, VA 22901, USA}

\end{center}

\renewcommand{\thefootnote}{\arabic{footnote}}
\setcounter{footnote}{0}
\thispagestyle{empty}



\begin{abstract}
We study the low energy implications of a trinification model based on the gauge symmetry $G= SU(3)_c \times SU(3)_L \times SU(3)_R$, without imposing gauge coupling unification. A minimal model requires two Higgs multiplets that reside in the bi-fundamental representation of $G$, and this is shown to be adequate for accommodating the Standard Model (SM) fermion masses and generate, via loop corrections and seesaw mechanism, suitable masses for the heavy neutral leptons as well as the observed SM neutrinos. We estimate a lower bound of around 15 TeV for the masses of the new down- type quarks that are required by the $SU(3)_L \times SU(3)_R$ symmetry.
We examine the resonant production at the LHC of the new gauge bosons, which leads to a lower bound of
16 TeV for the symmetry breaking scale of $G$. We also show how the muon $g-2$ anomaly can be resolved in the presence of these new gauge bosons and the heavy charged leptons present in the model.
Finally, the model predicts the presence of a topologically stable monopole carrying three quanta $(6 \pi /e)$ of Dirac magnetic charge and mass $\gtrsim 160$ TeV. If new matter fields lying in the fundamental representations of G are included, the model predicts the presence of exotic leptons, mesons and baryons carrying fractional electric charges such as $\pm e/3$ and $\pm 2e/3$, fully compatible with the Dirac quantization condition.

\end{abstract}

\newpage

{\hypersetup{linkcolor=black}
\tableofcontents}
\setcounter{footnote}{0}

\newpage
\section{Introduction}\label{SEC-01}

The trinification symmetry  $SU(3)_c \times SU(3)_L \times SU(3)_R$, without an $E_6$ embedding \cite{Gursey:1975ki, Shafi:1978gg, Achiman:1978vg} is arguably the simplest non-abelian gauge symmetry which can be safely broken to the SM gauge group, $SU(3)c \times SU(2)_L \times U(1)_Y$, at the TeV scale due to the absence of gauge-mediated proton decay \cite{Rujula:1984, Babu:1985gi,Dvali:1994vj,Dvali:1994wj,Kephart:2001ix, Willenbrock:2003ca, Kim:2004pe,Sayre:2006ma,Kephart:2006zd, Cauet:2010ng, Hetzel:2015bla, Hetzel:2015cca,Pelaggi:2015kna,Babu:2017xlu, Wang:2018yer, Babu:2021hef}. 
For comparison, the lower bound on the $SU(4)_c \times SU(2)_L \times SU(2)_R$ \cite{Pati:1974yy} symmetry breaking is around $10^3$ TeV \cite{Valencia:1994cj, Smirnov:2007hv}. 
In addition to charge quantization, trinified models also yeilds the desired value of the electroweak mixing angle with $g_R \simeq 0.71\;g_L$, where $g_L \simeq 0.63$ at ${\cal O} (10)$ TeV and $g_R$ denote the $SU(2)_L$ and $SU(2)_R$ gauge couplings, respectively. 

The main goal of this paper is to construct a realistic trinification model that can be realized near the TeV scale \cite{Pelaggi:2015kna}. Perhaps the most interesting aspects of such a TeV scale trinification model is that it offers a rich phenomenology accessible in collider experiments at the Large Hadron Collider (LHC) and its proposed upgrades. 
These prediction include twelve new gauge bosons and a number of charged leptons and quarks.
Interestingly, all the new gauge bosons masses are uniquely determined by just two parameters associated with the trinification breaking scale once the lightest gauge boson mass is fixed \cite{Babu:2017xlu}. A detailed study of the various phenomenological aspects of the trinification model, including a determination of the lower bound at which the trinification symmetry can be broken, is one of the goals of this work.

We show that two Higgs fields in the representation $(1,3,3^*)$ are sufficient to break the trinification symmetry down to $SU(3) \times U(1)_{EM}$ \cite{Babu:2017xlu} at the TeV scale, and this is consistent with realizing realistic masses for the SM fermions. Interestingly, the mass ratio between the lightest and the heaviest new down-type quarks can be around 10 or so. Setting the lightest new down-type quark mass to be $1.5$ TeV, the lower bound set from the LHC \cite{CMS:2020ttz}, the heaviest quarks mass must be at  least around $15$ TeV.
With three copies of Higgs fields, we find that the heavy quark masses are independent of the SM quark masses, and all the fermion masses and mixing can be easily accommodated, along with the TeV scale symmetry breaking.

We examine the resonance production of the new charged and neutral gauge bosons at the LHC. 
Since both $g_{L}$ and $g_R$ are fixed, the LHC upper bound on the cross section of a resonantly produced  gauge boson yields a lower bound on the new gauge boson masses, or equivalently, a lower bound on the trinification breaking scale $V \gtrsim 16$ TeV. 
This is comparable if not stronger than the bounds obtained from fermion mass fitting in the case with two copies of Higgs field. 

The SM singlet neutrinos present in the model are massless at the tree level, but we show how they acquire moderately large Majorana masses through one-loop radiative corrections, suppressed by a one-loop factor $(16 \pi^2)$, involving the scalars. We observe that these masses do not rely on the electroweak symmetry breaking parameters and scale linearly with the vacuum expectation values (VEVs) that break $SU(3)_L \times SU(3)_R$ down to the SM gauge group. The singlet neutrino in the theory gets its masses through the effective operator $\psi \psi  \Phi^\dagger \Phi^\dagger$, which sets an avenue to generate the observed light neutrinos masses and oscillations via the usual seesaw mechanism \cite{Weinberg:1979sa}. Note that the same operator can also induce small masses for the SM-like neutrinos through one-loop diagrams. Hence, the trinification model allows both type-I \cite{Minkowski:1977sc, Mohapatra:1979ia, Yanagida:1980xy, Gell-Mann:1979vob, Glashow:1979nm} and type-II \cite{Schechter:1980gr, Cheng:1980qt, Mohapatra:1980yp, Lazarides:1980nt} radiative seesaw scenarios.


A recent measurement of the muon anomalous magnetic moment $a_\mu$ at Fermilab National Accelerator Laboratory~(FNAL) reports $a_\mu(\textnormal{FNAL})=116592040(54)\times  10^{-11}$ ~\cite{Muong-2:2021ojo}, which agrees with the previous Brookhaven National Laboratory~(BNL) E82 measurement~\cite{Muong-2:2006rrc,Muong-2:2001kxu}. In contrast, a theoretical determination in the SM is reported to be $a_\mu^{\textnormal{SM}}=116591810(43)\times 10^{-11}$ \cite{Davier:2019can, Aoyama:2020ynm,Davier:2010nc,Gerardin:2020gpp}. The deviation, $\Delta a_\mu = a_\mu (\text{experiment})- a_\mu (\text{SM}) \simeq (251\pm59)\times 10^{-11}$ is a $4.2\sigma$ discrepancy. We show that trinification can eliminate this discrepancy with the correct sign through the exchange of a new neutral vector boson and muon-like heavy fermion at one-loop order. An upper bound of the relevant neutral gauge boson mass turns out to be about 9 TeV. 

The spontaneous breaking of $SU(3)_L \times SU(3)_R$ to $SU(2)_L \times U(1)_Y$ also predicts the presence of topologically stable magnetic monopole with three quanta of Dirac magnetic charge \cite{Shafi:1984wk,Lazarides:1986rt,Lazarides:1988wz,Kephart:2006zd,Kephart:2017esj, Lazarides:2019xai} and a mass about an order of magnitude larger than the symmetry breaking scale. 
Trinification also predicts the existence of exotic fractionally charged lepton (f-leptons), mesons and baryons. 
For monopole and exotic particle searches at the LHC, see Refs.~\cite{ MoEDAL:2021mpi, ATLAS:2019wkg, HassanElSawy:2744867,Acharya:2021ckc, CMS:2012xi,Pinfold:2019zwp,milliQan:2021lne}.

The rest of the paper is organized as follows. The trinification model with two Higgs multiplets is outlined in Sec. \ref{SEC-02}, which includes details of fermion mass generation, the scalar sector in the electroweak limit, and an analysis of the gauge boson masses and eigenstates. In Sec.~\ref{sec:collider} we discuss resonant production of the new gauge bosons at the LHC. Neutral lepton masses are considered in Sec.~\ref{sec:nuFermion}, and 
Sec.~\ref{sec:amm} shows how muon $g-2$ can be incorporated.
In Sec.~\ref{sec:monoploe} we discuss the topologically stable monopole as well as fractionally charged leptons, mesons and baryons, and our conclusions are summarized in Sec.~\ref{sec:conclusion}.

\section{Model}\label{SEC-02}
The model is based on the gauge symmetry $SU(3)_C \times SU(3)_L \times SU(3)_R$ under which the quarks ($Q_{L/R}$) and leptons ($\psi_L$) fields are represented as
\begin{equation}
    Q_{L}\left(3,3^{*}, 1\right)=\begin{pmatrix}
     u \\
d \\
D
\end{pmatrix}_L, 
\quad 
Q_{R} \left(3,1,3^{*}\right) =\begin{pmatrix}
 u \\
d \\
D
\end{pmatrix}_R \, ,
\quad 
\psi_{L}\left(1,3,3^{*}\right)= \begin{pmatrix}
 E^{0} & E^{-} & e^{-} \\
E^{+} & E^{c 0} & \nu \\
e^{c} & \nu^{c} & N
\end{pmatrix}_L \, , 
\label{eq:fermion}
\end{equation}
where $D_L$ and $D_R$ are new $SU(2)_L$ singlet quarks, ($E^0, E^+)$ and $(E^-, E^{c0}$) are new $SU(2)_L$ lepton doublets, and $\nu^c$ and N are SM singlet neutral leptons. Here $SU(3)_L$ acts vertically and $SU(3)_R$ acts horizontally. Note that lepton number is explicitly violated in the model because $e^c$ and $e^-$ reside in the same matter multiplet. 
The charge operator is given by 
\begin{eqnarray}
    Q = T_{3L} + T_{3R} + \frac{1}{\sqrt{3}} (T_{8L} + T_{8R}), 
    \label{eq:Qop}
\end{eqnarray}
where $T_{3} = (1/2)\;{\rm diag} (1,-1,0)$ and $T_{8} = (1/2\sqrt{3})\;{\rm diag} (1,1,-2)$ are the conventionally normalized $SU(3)_{L,R}$ generators. 

At least two Higgs fields $\Phi_{1,2}$ in the $(1, 3, 3^\star)$ representation are necessary in order to break the trinification symmetry down to $SU(3)_c \times U(1)_{EM}$ \cite{Pelaggi:2015kna} and to generate a realistic fermion mass spectrum. 
The vacuum expectation values (VEVs) for $\Phi_{1,2}$ are given by 
\begin{equation}
\left\langle\Phi_{1}\right\rangle=\left(\begin{array}{ccc}
v_{u1} & 0 & 0 \\
0 & v_{d1} & 0 \\
0 & 0 & V_1
\end{array}\right), \hspace{5mm}
\left\langle\Phi_{2}\right\rangle=\left(\begin{array}{ccc}
v_{u 2} & 0 & 0 \\
0 & v_{d 2} & v_{L 2} \\
0 & V_{R } & V_2
\end{array}\right), \hspace{5mm}
\label{eq:vev}
\end{equation}
where, for simplicity, all the VEVs are chosen to be real and the (3,2) and (2,3) elements of $\langle\Phi_1\rangle$ are set to zero by a gauge rotation. 
The VEVs $V_{1,2,R}$ break $SU(3)_L \times SU(3)_R$ to $SU(2)_L \times U(1)_{Y}$ which, in turn, is broken to $U(1)_{EM}$ by $v_{u1, u2}$, $v_{d1, d2}$, and $v_{L2}$. Individually, $V_{1,2}$ and $V_R$ can only break trinification to the left-right symmetry group $ SU(2)_L \times SU(2)_R \times U(1)_{B-L}$. 
The VEVs $v_{u1, u2}$, $v_{d1, d2}$, and $v_{L2}$ are of the order of electroweak scale, and  $V_{1,2,R}$ are at least of order TeV scale to be consistent with the LHC bounds on the heavy gauge boson masses \cite{ATLAS:2019fgd, ATLAS:2019erb}. 
The gauge boson mass spectrum will be discussed in detail in Sec.~\ref{sec:gauge}.

\subsection{Charged Fermion Masses}\label{sec:fermionmass}
There are 27 Weyl fermions per generation in the model of which 15 are the usual SM chiral fermions, as can be seen from Eq.~\eqref{eq:fermion}. The most general Yukawa interactions of quark and leptons with the Higgs fields $\Phi_{n= 1,2}$ are expressed as
\begin{equation}
    -{\cal L}_Y = Y_{qn}^{ab}\ \Bar{Q}_{L \alpha}^a (\Phi_{n})_i^\alpha Q_{R}^{bi} + Y_{L n}^{ab}\ \psi_i^{a,\alpha} \psi_j^{b,\beta} (\Phi_{n})_k^\gamma \epsilon^{ijk} \epsilon_{\alpha \beta \gamma} + h.c. \, . 
    \label{eq:YukawaQL}
\end{equation}
Here $(a,b)$ and $(i, j, k, \alpha, \beta, \gamma)$ are  respectively the generation and $SU(3)$ indices, and $Y_{qn, Ln}$ denote the $3\times 3$ Yukawa coupling matrices, and repeated indices imply summation.
From Eq.~\eqref{eq:YukawaQL} we obtain the following mass matrices for up-type and down-type quarks ($M_u$ and $M_d$) and charged leptons $(M_\ell)$:
\begin{equation}
M_{u}= Y_{q n}\ v_{u n}  \, , \hspace{10mm}
M_d = 
\begin{pmatrix}
 Y_{q n} v_{d n}  &~ Y_{q 2} V_R\\
  Y_{q 2} v_{L 2}  &~ Y_{q n} V_n 
\end{pmatrix}  \, , \hspace{10mm}
M_\ell = 
\begin{pmatrix}
 -Y_{L n} v_{d n}  &~ Y_{L 2} V_R\\
  Y_{L 2} v_{L 2}  &~ -Y_{L n} V_n
\end{pmatrix} \, .
  \label{eq:fermioncha}
\end{equation}
The basis for $M_d$ and $M_\ell$ are $(d, D)$ and  $(e, E)$, respectively. 
The SM singlet neutral fields $\nu^c$ and $N$ do not acquire mass at tree level from the Yukawa interaction of Eq.~\eqref{eq:YukawaQL}. 
Their masses are generated via quantum corrections at the one-loop level (a detailed analysis of the neutral fermion sector will be separately discussed in Sec.~\ref{sec:nuFermion}). 

Let us first examine the quark masses. It is essential to point out that setting $Y_{q2} \to 0$ decouples the SM quarks from the  heavy quarks. However, this choice is inconsistent as it results in the same mass hierarchy in the up and down-type quark sector across all three quark generations. With $Y_{q2} \neq 0$, the resulting down-type mass matrix can be block-diagonalized using a biunitary transformation, namely, 
\begin{equation}
    U_R M_d U_L^\dagger = 
    \begin{pmatrix}
     \hat{m}_d & 0 \\
     0 & \hat{m}_D
    \end{pmatrix} \, ,  
    \label{eq:downMat}
\end{equation}
where $U_{L,R}$ are unitary matrices and the $3\times 3$ light ($\hat{m}_d$) and heavy ($\hat{m}_D$) mass matrices are given by
\begin{align}
    \hat{m}_d &= \Big\{ Y_{q n}\ v_{d n} - \ V_R  v_{L 2}\ Y_{q2} (Y_{qn} V_{n})^{-1} Y_{q2} \Big\} \Big\{ I + Y_{q2} V_R (|Y_{qn} V_n|^2)^{-1} (Y_{q2} V_R)^\dagger \Big\}^{-1/2} \, , \nonumber \\
    \hat{m}_D &= \sqrt{|Y_{q 2} V_R|^2 + |Y_{q n} V_n|^2} \, .   
    \label{eq:dmass}
\end{align}
Here we have neglected corrections of order $\mathcal{O}(v_{ew}/V)$, with $v_{ew} = 246$ GeV and 
\begin{equation}
  V^2 = V_1^2 + V_2^2 + V_R^2.   
\end{equation}
The mass eigenstate are defined as 
\begin{equation}
     \begin{pmatrix}
     \hat{d}_L \\
     \hat{D}_L
    \end{pmatrix} = \begin{pmatrix}
     I & U_{L}^{12*}  \\[3pt]
    U_{L}^{21*} & I 
   \end{pmatrix} 
    \begin{pmatrix}
     d_L \\
     D_L
     \end{pmatrix} \, , \hspace{10mm}
    \begin{pmatrix}
     \hat{d}_R \\
     \hat{D}_R
    \end{pmatrix} = \begin{pmatrix}
    U_R^{11}  &  U_R^{12} \\[3pt]
      U_R^{21} &  U_R^{22}
    \end{pmatrix}^T 
    \begin{pmatrix}
     d_R \\
     D_R
    \end{pmatrix} \, .  
\end{equation}
The entries of unitary matrices that $U_{L,R}$ block-diagonalize the down-type matrix given in Eq.~\eqref{eq:downMat} are as follows: 
\begin{align}
  U_{L}^{12} &= -\Big[ (Y_{qn} v_{dn})^\dagger Y_{q2} V_R + (Y_{qn} v_{L2})^\dagger Y_{q2} V_n  \Big]\ \Big[ |Y_{q 2} V_R|^2 + |Y_{q n} V_n|^2 \Big]^{-1} \nonumber \, , \\
  U_L^{21} &= \Big[(Y_{q2} V_R)^\dagger Y_{qn} v_{dn} + (Y_{qn} V_n)^\dagger Y_{q2} v_{L2} \Big]  \Big[ |Y_{q 2} V_R|^2 + |Y_{q n} V_n|^2 \Big]^{-1} \nonumber \, ,\\
   U_R^{11} &= \Big[ I + Y_{q2} V_R (|Y_{qn} V_n|^2)^{-1} (Y_{q2}^\dagger V_R)^\dagger \Big]^{-1/2} \nonumber \, ,\\
  U_R^{12} &= -(Y_{q2} V_R) (Y_{qn} V_n) \Big[ I + Y_{q2} V_R (|Y_{qn} V_n|^2)^{-1} (Y_{q2}^\dagger V_R)^\dagger \Big]^{-1/2} \nonumber \, ,\\
  U_R^{21} &= (Y_{q 2} V_R)^\dagger \Big[|Y_{q 2} V_R|^2 + |Y_{q n} V_n|^2 \Big]^{-1/2} \nonumber \, ,\\
  U_R^{22} &= (Y_{q n} V_n)^\dagger \Big[|Y_{q 2} V_R|^2 + |Y_{q n} V_n|^2 \Big]^{-1/2} \, .
  \label{eq:MixULR}
\end{align}
From Eq.~\eqref{eq:MixULR} we find that the mixing between $d_L$ and $D_L$ parameterized by $U_L^{12}$ and $U_L^{21}$ depends on the electroweak symmetry breaking, and is of order $\mathcal{O} (v_{ew}/\hat{m}_D)$. However, the mixing between $d_R$ and $D_R$ does not require $SU(2)_L$ breaking, and so the mixing entries in $U_R^{12}$ and $U_R^{21}$ can be $\mathcal{O} (1)$. 

Analogous to the down-type quark matrix discussed above, charged lepton matrices in Eq.~\eqref{eq:fermioncha} can be block-diagonalized by substituting
\begin{align}
  Y_{q n}\ v_{d n}  &\to  -Y_{L n}\ v_{d n}  \, , \hspace{3mm}  Y_{q 2} V_R \to Y_{L 2} V_R \, , \hspace{3mm} Y_{q 2}\ v_{L 2}  \to Y_{L 2}\ v_{L 2} \, , \hspace{3mm} Y_{q n} V_n \to -Y_{L n} V_n \, ,
\end{align}
in the results for the down-type quark mass eigenvalues and eigenstates. Thus the lepton masses to the lowest order $\mathcal{O}(v_{ew}/V)$ are given by
\begin{align}
    \hat{m}_{e^-} &= \Big\{ -Y_{L n} v_{d n} + V_R  v_{L 2}\ Y_{L2} (Y_{Ln} V_{n})^{-1} Y_{L2} \Big\} \Big\{ I + Y_{L2} V_R (|Y_{Ln} V_n|^2)^{-1} (Y_{L2} V_R)^\dagger \Big\}^{-1/2} \, , \nonumber \\
    \hat{m}_{E^-} &= \sqrt{|Y_{L 2} V_R|^2 + |Y_{L n} V_n|^2} \, .
    \label{eq:dmass}
\end{align}
Analogous to the $d_R-D_R$ mixing case, the mixing between $e^-$ and $E^-$ can be significant. 


\subsection{Trinification Breaking Scale from Fermion Mass Fitting }

As mentioned earlier, the up-type, down-type, and the new heavy quarks have the same flavor structure with just one copy of $\Phi_n$ which, of course, is unacceptable. Furthermore, the need for at least two copies of $\Phi$ is obvious; with only one $\Phi$, its diagonalized VEV will preserve an unbroken $SU(2)_L \times SU(2)_R \times U(1)$ symmetry. 
In this section, we will determine the lowest trinification breaking scale by minimizing the mass ratio between the heaviest and the lightest exotic quark. 
To simplify the analysis, let us consider small mixing between $d_R$ and $D_R$ by taking $Y_{q2} \ll \hat{m}_D$ in Eq.~\eqref{eq:MixULR}. Focusing only on the SM up and down-type quarks allows us to invert the mass matrix equations and obtain
\begin{equation}
    Y_{q1} \simeq \frac{{M}_u v_{d2} - {\tilde M}_d  v_{u2}}{v_{u1} v_{d2} - v_{u2} v_{d1}} \, ,
    \hspace{10mm} Y_{q2} \simeq \frac{- { M}_u v_{d1} + {\tilde M}_d  v_{u1}}{v_{u1} v_{d2} - v_{u2} v_{d1}} \, ,
    \label{eq:Yq12}
\end{equation}
where ${\tilde M}_{d}$ is the light down-type type $3\times3$ mass matrix in the limit $Y_{q2} \ll \hat{m}_D$.
We further require $v_{u1} v_{d2} \neq v_{u2} v_{d1}$, for otherwise we obtain ${\tilde M}_d = (v_{d2}/v_{u2}){ M}_u$, which is inconsistent with observations. 
Choosing a diagonal basis for the up-type quarks, the down-type mass matrix is expressed as 
\begin{equation}
    {\tilde M}_d = V { M}_d^{\text{diag}} V'^\dagger\, . 
\end{equation}
Here $V'$ is an arbitary unitary matrix parameterized as $d_R = V d_R^0$ ($d_R^0$ is the mass eigenstates vector), and $V$ is related to the Cabibbo-Kobayashi-Maskawa (CKM) quark mixing matrix $V_{\rm CKM}$ and the diagonal phase matrices $P$ and $Q$ as 
\begin{equation}
   V = P\; V_{\rm CKM} \;Q \, .   
\end{equation}
Together with Eq.~(\ref{eq:Yq12}), the $3\times 3$ heavy-quark mass matrix  can be expressed as
\begin{equation}
    {\tilde M}_D =  a \left( M_u^{\rm diag} + b\ P\ V_{\rm CKM}\  Q\ M_d^{\rm diag}\ V'^\dagger  \right) \, ,
    \label{eq:exoticD}
\end{equation}
where ${M}_u^{\rm diag}$ and ${M}_d^{\rm diag}$ are diagonal matrices corresponding to SM up and down-type quarks, respectively, and $a$ and $b$ are defined as
\begin{equation}
    a =  \frac{V_1 v_{d2} - V_2 v_{d1}}{v_{u1} v_{d2} - v_{u2} v_{d1}} \, , \hspace{10mm} b = \frac{V_1 v_{u2} - V_2 v_{u1}}{v_{u1} v_{d2} - v_{u2} v_{d1}} \, .
\end{equation}

We have numerically diagonalized the heavy quark mass matrix in Eq.~\eqref{eq:exoticD} to find the minimum ratio between the heaviest and lightest eigenvalues. 
We perform a parameter scan over the free parameters, namely, three angles from $V'$, 8 phases and $b$. 
We find the minimum value for the mass ratio $m_{D_3}/m_{D_1} \simeq 10$. 
The mass of the lightest new down-type quark with hypercharge $(-1/3)$ is bounded from below by the LHC at around $1.5$ TeV \cite{CMS:2020ttz}. 
Identifying this with $m_{D1}$ and together with the parameter scan results, we obtain a lower bound on $m_{D_3} \simeq 15$ TeV. 
This bound can be approximately interpreted as a lower bound on the trinification symmetry breaking scale $V$. 

Before concluding this section we briefly discuss the case with three copies of $\Phi_n$. 
The masses of $M_u$, $M_d$, and $M_D$ are easily compatible with the experiments using three Yukawa couplings $Y_{qn} (n=1-3)$. 
However, we lose the dependence of heavy quark masses on the SM quark masses. 
To illustrate this, let us consider the case in which all the electroweak scale VEVs in $\Phi_1$ are set to zero, \textbf{$v_{u1}= v_{d1} = v_{L1} = 0$}, with $\Phi_1$ breaking $SU(3)_L \times SU(3)_R$ to $SU(2)_L \times SU(2)_R \times U(1)_{B-L}$, which is then broken to $U(1)_{em}$ by $\Phi_2$ and $\Phi_3$. 
This leads to the following mass relations:
\begin{equation}
     M_D = Y_{q1} V_1 \, , \hspace{3mm} M_{E^-} = Y_{L1} V_1 \, , \hspace{3mm} 
    M_u = Y_{qm} v_{um} \, , \hspace{3mm}  M_d = Y_{qm} v_{dm} \, ,  \hspace{3mm}  M_{e^-} = Y_{Lm} v_{dm} \, ,
\end{equation}
where the repeated indices $m = 2,3$ are summed over. 
It is clear that the SM up- and down-type quark masses are independent of the heavy quark masses. Requiring the heavy fermion masses to be at the TeV scale, we find that the trinification symmetry can be broken at a few TeV with order one values for $Y_{q1}$ and $Y_{L1}$. 
As we discuss later in Sec.~\ref{sec:gauge}, in this case the most stringent bound on the trinification symmetry breaking scale is from gauge boson searches which directly constrain the VEVs.

\subsection{Scalar Sector in the Electroweak Limit}
In this section we construct and analyze the Higgs potential with two copies of $(1,3,3^*)$ with $\Phi_i^\alpha$, where $i$ and $\alpha$ are respectively the $SU(3)_L$ and $SU(3)_R$ indices. The most renormalizable Higgs potential with $\Phi_{1,2}$ is given by
\begin{align}
    V_{POT} =&\ m_1^2\ Tr(\Phi_1^\dagger \Phi_1) + m_2^2\ Tr(\Phi_2^\dagger \Phi_2) + m_{12}^2\ \{ Tr(\Phi_1^\dagger \Phi_2) + Tr(\Phi_1 \Phi_2^\dagger) \}  \nonumber \\
    & + \mu_1 \Phi_{1i}^\alpha \Phi_{1j}^\beta \Phi_{1k}^\gamma \epsilon^{ijk} \epsilon_{\alpha \beta \gamma} + \mu_2 \Phi_{2i}^\alpha \Phi_{2j}^\beta \Phi_{2k}^\gamma \epsilon^{ijk} \epsilon_{\alpha \beta \gamma}  + \mu_3 \Phi_{1i}^\alpha \Phi_{1j}^\beta \Phi_{2k}^\gamma \epsilon^{ijk} \epsilon_{\alpha \beta \gamma}  \nonumber \\
    & + \mu_4 \Phi_{1i}^\alpha \Phi_{2j}^\beta \Phi_{2k}^\gamma \epsilon^{ijk} \epsilon_{\alpha \beta \gamma} + \lambda_1 Tr(\Phi_1^\dagger \Phi_1)^2 + \lambda_2 Tr(\Phi_1^\dagger \Phi_1 \Phi_1^\dagger \Phi_1) + \lambda_3 Tr(\Phi_2^\dagger \Phi_2)^2  \nonumber \\
    & + \lambda_4 Tr(\Phi_2^\dagger \Phi_2 \Phi_2^\dagger \Phi_2)+ \lambda_5 Tr(\Phi_1^\dagger \Phi_1) Tr(\Phi_2^\dagger \Phi_2) + \lambda_6 Tr(\Phi_1^\dagger \Phi_1 \Phi_2^\dagger \Phi_2)  \nonumber \\
    & + \lambda_7 Tr(\Phi_1^\dagger \Phi_2) Tr(\Phi_1 \Phi_2^\dagger) \nonumber + \lambda_8 Tr(\Phi_1^\dagger \Phi_2 \Phi_2^\dagger \Phi_1) +  \lambda_9 Tr(\Phi_1^\dagger \Phi_2)^2 + \lambda_{10} Tr(\Phi_1^\dagger \Phi_2 \Phi_1^\dagger \Phi_2)  \nonumber \\
    & + \lambda_{11} Tr(\Phi_1^\dagger \Phi_1) Tr(\Phi_1^\dagger \Phi_2) + \lambda_{12} Tr(\Phi_1^\dagger \Phi_1 \Phi_1^\dagger \Phi_2) + \lambda_{13} Tr(\Phi_2^\dagger \Phi_2) Tr(\Phi_1^\dagger \Phi_2) 
    \nonumber \\
    & + \lambda_{14}  Tr(\Phi_2^\dagger \Phi_2 \Phi_1^\dagger \Phi_2) + h.c. , 
    \label{eq:pot}
\end{align}
where all of the parameters including the VEVs are taken to be real for simplicity. The mass matrices for the charged and neutral scalar fields can be constructed by inserting the VEVs of Eq.~\eqref{eq:vev} in Eq.~\eqref{eq:pot}. 

A complete analysis of the Higgs potential including the electroweak VEVs in $\Phi_{1,2}$ is beyond the scope of the current work. Instead, we analyze the Higgs potential in the electroweak conserving limit by setting the electroweak VEVs $v_{un} = v_{dn} = v_{Ln} = 0$. We show the consistency of symmetry breaking by properly identifying the 12 Goldstone modes associated with the spontaneous breaking of $SU(3)_L \times SU(3)_R$ down to the SM gauge symmetry. 
The stationary conditions to minimize the Higgs potential, 
\begin{equation}
    \frac{\partial V_{POT}}{\partial V_{1}} = 0,  \qquad
    \frac{\partial V_{POT}}{\partial V_{2}} = 0, \qquad
    \frac{\partial V_{POT}}{\partial V_{R}} = 0, 
\end{equation} 
yield the following relations: 
\begin{align}
    m_1^2 &= -2(\lambda_1+ \lambda_2) V_1^2 - (\lambda_5 + \lambda_8) (V_2^2 + V_R^2) - 2 (\lambda_{11} + \lambda_{12}) V_1 V_2 \, , \nonumber \\
    m_2^2 &= -(\lambda_5+ \lambda_8) V_1^2 - 2(\lambda_3 + \lambda_4) (V_2^2 + V_R^2) - 2 (\lambda_{13} + \lambda_{14}) V_1 V_2 \, , \nonumber \\
    m_{12}^2 &= -(\lambda_{11}+ \lambda_{12}) V_1^2 - (\lambda_{13} + \lambda_{14}) (V_2^2 + V_R^2) -  (2\lambda_{10} + 2\lambda_{9} + \lambda_6 + \lambda_7) V_1 V_2 \, .
    \label{eq:minCond}
\end{align}

Substituting Eq.~\eqref{eq:minCond} in Eq.~\eqref{eq:pot} we obtain the mass matrices for the charged and neutral components of the Higgs fields and identify the Goldstone modes which are eigenstates corresponding to null eigenvalues. 
In the electroweak conserving limit, the $SU(2)_L$ doublet fields $ \{ (\phi_1^\alpha)_n$, $(\phi_2^\alpha)_n \}$ do not mix with $SU(2)_L$ singlet fields $(\phi_3^\alpha)_n$. The real and pseudoscalar components of the neutral Higgs fields do not mix as all the parameters are taken to be real. Thus, in this limit, the doublet scalar sector mass matrices for the charged, real and pseudoscalar fields have an identical $6\times 6$ mass matrix $M$, which is diagonalized by a single unitary matrix, U,
\begin{equation}
    U^T M U = M_{\text{diag}} \, ,
\end{equation}
where $M_{\text{diag}}$ is a diagonal mass matrix in the mass basis. This mass matrix $M$ is real and symmetric $M_{ij} = M_{ji}$, with
\begin{equation}
    \begin{array}{ll}
       M_{11} = - 2 \lambda_2 V_1^2 - 2 \lambda_{12} V_1 V_2 - \lambda_8 (V_2^2 + V_R^2) \, ,  &~~~ M_{12} =-\mu_1 V_1 - \mu_3 V_2   \, , \\
      M_{13} = \mu_3 V_R \, ,  &~~~ M_{14} =-\lambda_{12} V_1^2 - \lambda' V_1 V_2 -\lambda_{14} (V_2^2 + V_R^2) \, , \\
        M_{15}  = M_{24} =  -\mu_3 V_1 - \mu_4 V_2 \, ,  &~~~   M_{16} = M_{34} =  \mu_4 V_R \, , \\
      M_{22} = M_{11} - \lambda_6 V_R^2 \, ,  &~~~ M_{23} =  \lambda_{12} V_1 V_R + \lambda_6 V_2 V_R \, , \\
       M_{25} = M_{14} + \lambda_{14} V_R^2 \, ,  &~~~  M_{26} = 2 \lambda_{10} V_1 V_R + \lambda_{14} V_2 V_R \, , \\
      M_{33} = \lambda''  V_2^2 - \lambda_8 V_R^2 \, ,  &~~~ M_{35} = \lambda_8 V_1 V_R + \lambda_{14} V_2 V_R \, , \\
       M_{36} = -\lambda''  V_1 V_2 - \lambda_{14} V_R^2 \, ,  &~~~ M_{44} = -\lambda_8 V_1^2 - 2 \lambda_{14} V_1 V_2 - 2 \lambda_4 V_2^2 - 2 \lambda_4 V_R^2 \, , \\
       M_{45} =-\mu_4 V_1 - \mu_2 V_2 \, ,  &~~~  M_{46} = \mu_2 V_R \, , \\
        M_{55} = M_{44} + 2 \lambda_4 V_2^2 \, ,  &~~~  M_{56} = \lambda_{14} V_1 V_R + 2 \lambda_4 V_2 V_R \, , \\
       M_{66} =  (\lambda_{6}- \lambda_8) V_1^2 - 2 \lambda_4 V_R^2 \, .  &
    \end{array}
    \label{eq:mscalar}
\end{equation}
where $\lambda' = 2 \lambda_{10} + \lambda_6$ and $\lambda'' = \lambda_6 - \lambda_8$. 

To identify the Goldstone modes and the physical massive scalar states, we decompose the neutral scalar into its real ($r$) and pseudoscalar ($i$) components, for instance, $\phi_1^1 = 1/\sqrt{2}\ (\phi_1^{1r} + i \phi_1^{1i}  )$. 
Using $U$, the flavor states for the charged and neutral scalars can be expressed in terms of the mass eigenstates as 
\begin{align}
    \left\{(\phi_2^1)_1, (\phi_1^{2*})_1, (\phi_1^{3*})_1, (\phi_2^1)_2, (\phi_1^{2*})_2, (\phi_1^{3*})_2 \right\}^T &= U\ \left\{G^+, h_1^+,  h_2^+,  h_3^+,  h_4^+,  h_5^+ \right\} ^T \nonumber \\
     \left\{(\phi_1^1)_1^r, (\phi_2^{2})_1^r, (\phi_2^{3})_1^r, (\phi_1^1)_2^r, (\phi_2^{2})_2^r, (\phi_2^{3})_2^r \right\}^T &= U\ \left\{G^{0r}, h_1^{0r},  h_2^{0r},  h_3^{0r},  h_4^{0r},  h_5^{0r} \right\} ^T \nonumber \\
     \left\{(\phi_1^1)_1^i, (\phi_2^{2})_1^i, (\phi_2^{3})_1^i, (\phi_1^1)_2^i, (\phi_2^{2})_2^i, (\phi_2^{3})_2^i \right\}^T &= U\ \left\{G^{0i}, h_1^{0i},  h_2^{0i},  h_3^{0i},  h_4^{0i},  h_5^{0i} \right\} ^T,    
    \label{eq:gold1}
\end{align}
where $h_\alpha^{+(0)}$ denote the physical  charged (neutral) scalar mass eigenstates and $(G^\pm, G^{0r}, G^{0i} )$ are the Goldstone modes. 
The latter modes are linear combination of $(\phi_1^3)_1, (\phi_1^3)_2,$ and $(\phi_1^2)_2,$   \begin{align}
    G^+ &= \frac{V_1 (\phi_1^{3*})_1 + V_2 (\phi_1^{3*})_2 + V_R (\phi_1^{2*})_2 
    }{\sqrt{V_1^2+V_2^2+V_R^2}} \nonumber \, ,\\ 
    G^{0r} &= \frac{V_1 (\phi_2^3)_1^r + V_2 (\phi_2^3)_2^r + V_R (\phi_2^2)_2^r 
    }{\sqrt{V_1^2+V_2^2+V_R^2}} \nonumber \, , \\
    G^{0i} &= \frac{V_1 (\phi_2^3)_1^i + V_2 (\phi_2^3)_2^i - V_R (\phi_2^2)_2^i 
    }{\sqrt{V_1^2+V_2^2+V_R^2}} \, ,
\end{align}
and $G^-$ is obtained by conjugation of $G^+$. 

The $SU(2)_L$ singlet sector includes a total of $8$ Goldstone modes, with the charged fields $(\phi_3^1)_1$ and $(\phi_3^1)_2$ both identified as Goldstone modes. 
The remaining four Goldstone modes reside in the neutral scalar fields $(\phi_3^2)_n$ and $(\phi_3^3)_n$. 
The real and symmetric mass matrix in the basis $\{(\phi_3^2)_1^r, (\phi_3^3)_1^r, (\phi_3^2)_2^r, (\phi_3^3)_2^r  \}$ read as follows:
\begin{equation}
    \begin{array}{ll}
        \tilde{M}_{11} = \tilde{\lambda} V_R^2  \, , &~~~  \tilde{M}_{12} = 2(\tilde{\lambda}' V_1 + \tilde{\lambda} V_2) V_R  \, ,\\
        \tilde{M}_{13} = 2 \tilde{\lambda}'' V_R^2  \, , &~~~ \tilde{M}_{14} = (\tilde{\lambda} V_1 + 2 \tilde{\lambda}'' V_2 ) V_R \, , \\
        \tilde{M}_{22} = 4 (\lambda_1+ \lambda_2) V_1^2 + 4 \tilde{\lambda}' V_1 V_2 + \tilde{\lambda} V_2^2 \, , &~~~ \tilde{M}_{23} = 2 ( (\lambda_5 + \lambda_8) V_1 + \tilde{\lambda}'' V_2 ) V_r \,  , \\
       \tilde{M}_{24} = 2 \tilde{\lambda}' V_1^2 + (\tilde{\lambda} + 2 \lambda_5 + 2 \lambda_8) V_1 V_2 + 2 \tilde{\lambda}'' V_2^2 \, ,  &~~~ \tilde{M}_{33} = 4 (\lambda_3 + \lambda_4) V_R^2 \, , \\
       \tilde{M}_{34} = 2(\tilde{\lambda}'' V_1 + 2 (\lambda_3 + \lambda_4) V_2) V_R \, , &~~~ \tilde{M}_{44} = \tilde{\lambda} V_1^2 + 4 V_2 ( \tilde{\lambda}'' V_1 + (\lambda_3 + \lambda_4) V_2 ) \, ,
    \end{array}
\end{equation}
where $\tilde{\lambda} = \lambda_6 + \lambda_7+ 2 \lambda_{10} + 2 \lambda_9$, $\tilde{\lambda}' = \lambda_{11}+ \lambda_{12}$, and $\tilde{\lambda}'' = \lambda_{13}+ \lambda_{14}$. The above matrix can be diagonalized numerically. The flavor states can be expressed in terms of the mass eigenstates by a $4\times 4$ unitary transformation matrix $U^\prime$, 
\begin{equation}
    \{(\phi_3^2)_1^r, (\phi_3^3)_1^r, (\phi_3^2)_2^r, (\phi_3^3)_2^r  \}^T = U'\ \{ G'^{0r}, h_1^{'0r}, h_2^{'0r}, h_3^{'0r} \}^T \, ,
\end{equation}
where $h_\alpha^{'0r}$ are the physical states and $G'^{0r}$ is a Goldstone mode identified as
\begin{equation}
    G^{'0r} = \frac{V_1 (\phi_3^2)_1^r + V_2 (\phi_3^2)_2^r - V_R (\phi_3^3)_2^r}{\sqrt{V_1^2 + V_2^2 + V_R^2}}
\end{equation}

Among the pseudoscalars in $(\phi_3^2)_n$ and $(\phi_3^3)_n$, one of the Goldstone modes is identified to be $(\phi_3^2)_2^i$. 
The remaining two Goldstone modes involve the pseudoscalars in $(\phi_3^2)_1^i$ and $(\phi_3^3)_n^i$.  These real and symmetric mass matrices in the basis $\{ (\phi_3^3)_1^i, (\phi_3^3)_2^i, (\phi_3^2)_1^i\}$ are expressed as
\begin{equation}
    \begin{array}{lll}
      \hat{M}_{11} = \hat{\lambda} V_R^2 \, ,  &~~~~~ \hat{M}_{12} = \hat{\lambda} V_2 V_R &~~~~~ \hat{M}_{13} = -\hat{\lambda} V_1 V_R \\
       \hat{M}_{22} = \hat{\lambda} V_2^2  &~~~~~ \hat{M}_{23} = -\hat{\lambda} V_1 V_2  &~~~~~ \hat{M}_{33} = \hat{\lambda} V_1^2 \, ,
    \end{array}
\end{equation}
where $\hat{\lambda} = \lambda_6 + \lambda_7 - 2 (\lambda_{9} + \lambda_{10})$. The two Goldstone modes are identified as
\begin{align}
    G_1^{'0i} &= \frac{V_1 (\phi_3^2)_1^i + V_R (\phi_3^3)_2^i }{\sqrt{V_1^2 + V_R^2}} \, ,\nonumber \\
    G_2^{'0i} &= \frac{-V_2 (\phi_3^2)_1^i + V_R (\phi_3^3)_1^i }{\sqrt{V_2^2 + V_R^2}} \, .
\end{align}
The massive physical state is, of course, simply the orthogonal state to the massless modes given above. Thus, we have identified a total of $12$ Goldstone modes which are absorbed by the $12$ heavy gauge bosons we expect after the trinification symmetry breaking, as we discuss in the next section.

\subsection{Gauge Sector}\label{sec:gauge}
In this section we analyze the gauge boson sector to evaluate all the masses and eigenstates. In particular, we find that the 12 new gauge bosons masses are uniquely determined in terms of the VEV ratio $V_R/V$ and the mass of the lightest new gauge boson, and our results are in agreement with Ref.~\cite{Babu:2017xlu}. 
Under the $SU(3)_L \times SU(3)_R$ symmetry, $\Phi_n \to U_L \Phi_n U_R^\dagger$, and the gauge kinetic term is given by 
\begin{equation}
    {\cal L}_{gauge} = \sum_n D_\mu (\Phi_n)_i^\alpha\ D^\mu (\Phi_n)_\alpha^i \, , 
    \label{eq:GaugeL}
\end{equation}
where the covariant derivative reads 
\begin{align}
    D^{\mu} (\Phi_{n})_i^\alpha &= \partial^{\mu} (\Phi_{n})_i^\alpha-\frac{i g_{L}}{2}\ (\vec{T} \cdot \vec{W}_{L}^{\mu})_i^k (\Phi_{n})_k^\alpha +\frac{i g_{R}}{2}\  (\vec{T} \cdot \vec{W}_{R}^{\mu})_k^\alpha\ (\Phi_{n})^k_i
    \, . 
\end{align}
Here $g_{L,R}$ are the $SU(3)_{L,R}$ gauge couplings, respectively, and the gauge boson multiplets in the $(1,8,1)_L$ and $(1,1,8)_R$ representation are defined as
\begin{equation}
\vec{T} \cdot \vec{W}_{L, R}^{\mu} =\left(\begin{array}{ccc}
W_3^{\mu} + \frac{W_8^{\mu}}{\sqrt{3}} & \sqrt{2} W^{+\mu} & \sqrt{2} V^{+\mu} \\
\sqrt{2} W^{-\mu} &- W_3^{\mu} + \frac{W_8^{\mu}}{\sqrt{3}}  & \sqrt{2} V^{0\mu } \\
\sqrt{2} V^{-\mu} & \sqrt{2}{V^{0\mu }}^{*} & -\frac{W_8^{\mu}}{\sqrt{3}}  \\
\end{array}\right)_{L,R} \, ,
\label{eq:cov}
\end{equation}
with the following definitions
\begin{equation}
    W^{\pm \mu} = \frac{W_1^\mu \mp i W_2^\mu}{\sqrt{2}}, \hspace{5mm} V^{\pm \mu} = \frac{W_4^\mu \mp i W_5^\mu}{\sqrt{2}}, \hspace{5mm}  V^{0(*)\mu} = \frac{W_6^\mu \mp i W_7^\mu}{\sqrt{2}}. 
    \label{eq:gaugeV}
\end{equation}
The new gauge bosons $(V^{+\mu}, V^{0\mu})_{L,R}$ are $SU(2)_{L,R}$ doublets, and $(W_8^\mu)_{L,R}$ are $SU(2)_{L,R}$ singlets. The gauge boson mass matrix is obtained by replacing the $\Phi$ fields with their VEVs in Eq.~\eqref{eq:GaugeL}. We adopt the parameterization given in Ref.~\cite{Babu:2017xlu} for convenience, namely,
\begin{align}
V^2 &=  V_R^2 + \sum_{n} V_n^2
\, , \hspace{4mm} 
S_{uu} = \sum_n  v_{un}^2
\, , \hspace{4mm} 
S_{dd} = \sum_n  v_{dn}^2
\, , \hspace{4mm} 
S_{ud} = \sum_n  v_{un} v_{dn}
\, , \hspace{4mm} 
S_{uL} =  v_{u2} v_{L2} \, ,
\nonumber \\ 
S_{uV} &= \sum_{n}  v_{un} V_n 
\, , \hspace{4mm}
S_{LV} =  v_{L2} V_2 
\, , \hspace{4mm}
S_{dL} =  v_{d2} v_{L2}
\, , \hspace{4mm} 
S_{dR} = v_{d2} V_R
\, , \hspace{4mm} 
S_{uR} = v_{u2} V_R\, .
\end{align}
For the charged gauge boson sector in the basis $(W_L^{\mu+}, V_L^{\mu+}, W_R^{\mu+}, V_R^{\mu+})$, the symmetric mass matrix is as follows:
\begin{equation}
    M_+^2 = 
    \begin{pmatrix}
     \frac{g_L^2}{2} (S_{dd} \!+\! S_{uu} \!+\! S_{LL}) & \frac{g_L^2}{2} (S_{LV} \!+\! S_{dR})  & - g_L g_R  S_{ud} &  -g_L g_R S_{uL} \\
     & \frac{g_L^2}{2} (S_{uu} \!+\! V^2)& -g_L g_R S_{uR} & -g_L g_R S_{uV} \\
     & & \frac{g_R^2}{2} (S_{dd} \!+\! S_{uu} \!+\! V_R^2) & \frac{g_R^2}{2} (S_{dL} \!+\! V_2 V_R) \\
     & & & \frac{g_R^2}{2}(S_{LL} \!+\! S_{uu} \!+\! V_1^2 \!+\! V_2^2) 
    \end{pmatrix}\; .
    \label{eq:charM}
\end{equation}
For $V_n, V_R >> v_{un}, v_{dn}, v_{Ln}$, the  mixing of $W_L^{\pm\mu}$ and $V_L^{\pm\mu}$ with $W_R^{\pm\mu}$ and $V_R^{\pm\mu}$ is of order $(v_{ew}/V)$, which is small. 
The mixing between $W_L^{\pm\mu}$ and $V_L^{\pm\mu}$ is also of order $(v_{ew}/V)$. Moreover, the mixing between $W_L^{\pm \mu}$ and $W_R^{\pm \mu}$ is strongly constrained to be $\leq 4 \times 10^{-3}$ from strangeness changing nonleptonic decays of hadrons \cite{Donoghue:1982mx}, as well as $b \to s \gamma$ \cite{Babu:1993hx}.
To the lowest order in $(v_{ew}/V)$, we identify the mass eigenstates as follows:
\clearpage
\begin{align}
    W_{1}^{\pm \mu} &\equiv W_{L}^{\pm \mu} \, , \nonumber \\
    W_{2}^{\pm \mu} &= \cos\theta_+ W_{R}^{\pm \mu} + \sin\theta_+ V_{R}^{\pm \mu}\, , \nonumber \\
    W_{3}^{\pm \mu} &= -\sin\theta_+ W_{R}^{\pm \mu} + \cos\theta_+ V_{R}^{\pm \mu} \, , \nonumber \\
    W_{4}^{\pm \mu} &\equiv V_{L}^{\pm \mu} \, ,
    \label{eq:gbC1}
\end{align}
where the mixing angle $\theta_+$ is given by
\begin{equation}
    \tan2\theta_+ \simeq \frac{-2 V_2 V_R}{V^2-2V_R^2} \, .
    \label{eq:gbC2}
\end{equation}
The mass eigenvalues to order in $\mathcal{O}(v_{ew}/V)$ are as follows:
\begin{align}
    m_{W_1}^{2} &=\frac{g_{L}^{2}}{2}\left(S_{u u}+S_{d d}+S_{L L}-\frac{\left(S_{LV}+S_{dR}\right)^{2}} {V^{2}}\right) \, ,\nonumber \\ 
    m_{W_{2}}^{2} &=\frac{g_{R}^{2} }{4} \left(V^{2} - \frac{V^2-2 V_R^2}{\cos2\theta_+}\right) \, ,\nonumber\\ 
    m_{W_{3}}^{2} &=\frac{g_{R}^{2} }{4} \left(V^{2} + \frac{V^2-2 V_R^2}{\cos2\theta_+}\right)  \, ,\nonumber\\ 
    m_{W_{4}}^{2} &=\frac{g_{L}^{2}}{2} V^{2} \, .
    \label{eq:GBC}
\end{align}
Here we identify $W_1^{\pm \mu}$ with mass proportional to the electroweak scale as the SM charged gauge boson $W^{\pm \mu}$. 
Requiring both $m_{W_2}^2$ and $m_{W_3}^2$ to be positive in Eq.~(\ref{eq:GBC}), we obtain the upper bound $V_2/V < \sqrt{1 - (V_R/V)^2}$. 


In the neutral sector, the flavor states $( W_{3L}^\mu, W_{3R}^\mu, W_{8L}^\mu, W_{8R}^\mu, W_{6L}^\mu, W_{6R}^\mu  )$ do not mix with the flavor states $( W_{7L}^\mu, W_{7R}^\mu )$, where $(W_{7}^\mu)_{L,R}$ are the imaginary components of the gauge field $V^{0\mu}$. 
The $6\times 6$ symmetric matrix $M_0$ spanning $( W_{3L}^\mu, W_{3R}^\mu, W_{8L}^\mu, W_{8R}^\mu, W_{6L}^\mu, W_{6R}^\mu  )$ has matrix elements that read
\begin{equation}
    \begin{array}{ll}
       (M_0)_{11} = \frac{g_{L}^{2}}{2} \left(s_{d d}+S_{L L}+S_{uu}\right) \, ,  &~~~~ (M_0)_{33} = \frac{g_{L}^2}{6} \left(S_{dd}+ S_{LL} + S_{uu} + 4 V^2 \right) \, , \\
       (M_0)_{12} = -\frac{g_{L} g_{R}}{2} \left(S_{d d}+S_{u u}\right) \, , &~~~~ (M_0)_{34} = -\frac{g_{L} g_R}{6} \left(S_{dd}-2 S_{LL} + S_{uu} + 4 V^2 - 6 V_R^2 \right) \, ,\\
       (M_0)_{13} = \frac{g_{L}^{2}}{2 \sqrt{3}}\left(S_{u u}-S_{d d}-S_{L L}\right) \, , &~~~~ (M_0)_{35} = -\frac{g_{L}^2}{2\sqrt{3}} \left(S_{LV}+ S_{dR}\right) \, , \\
       (M_0)_{14} = \frac{g_{L}g_R}{2 \sqrt{3}}\left(S_{dd}-2S_{LL}-S_{uu}\right) \, , &~~~~  (M_0)_{36} = \frac{g_{L} g_R}{2\sqrt{3}} \left(-S_{dL}+ 2 V_2 V_R\right) \, , \\
       (M_0)_{15} = -\frac{g_{L}^2}{2} \left(S_{LV}+S_{dR}\right) \, , &~~~~ (M_0)_{44} = \frac{g_R^2}{6} \left(S_{dd}+ 4 S_{LL} + S_{uu} + 4 V^2 - 2 V_R^2 \right) \\
       (M_0)_{16} = g_L g_R S_{dL} \, , &~~~~  (M_0)_{45} = \frac{g_{L} g_R}{\sqrt{3}} \left(2S_{LV} - S_{dR}\right)\, , \\
       (M_0)_{22} = \frac{g_{R}^2}{2} \left(S_{dd}+S_{uu}+V_R^2\right)  \, , &~~~~ (M_0)_{46} = -\frac{g_R^2}{2\sqrt{3}} \left(S_{dL} + V_2 V_R\right) \, , \\
       (M_0)_{23} = \frac{g_L g_R}{2 \sqrt{3}} \left(S_{dd}-S_{uu}-2V_R^2\right) \, , &~~~~ (M_0)_{55} = \frac{g_{L}^2}{2} \left(S_{dd} + S_{LL} + V^2 \right) \, ,\\
       (M_0)_{24} = -\frac{g_{R}^2}{2 \sqrt{3}} \left(S_{dd}-S_{uu}+V_R^2\right) \, , &~~~~ (M_0)_{56} = - g_{L} g_R \left(S_{dV} + S_{LR}\right) \, ,\\
       (M_0)_{25} = g_L g_R S_{dR} \, , &~~~~ (M_0)_{66} = \frac{g_R^2}{2} \left(S_{dd} + S_{LL} + V^2\right) \, ,\\
       (M_0)_{26} = -\frac{g_{R}^2}{2} \left(S_{dL}+ V_2 V_R \right) \, .& 
    \end{array}
    \label{eq:NeuM}
\end{equation}
The mass matrix $M_0$ includes a single massless state which is identified to the SM photon. 
A convenient basis to identify the eigenvectors and eigenvalues of $M_0$ is expressed as  
\begin{align}
    A^{\mu} &= \frac{\sqrt{3} g_{R} W_{3 L}^{\mu}+\sqrt{3} g_{L} W_{3 R}^{\mu}+g_{R} W_{8 L}^{\mu}+g_{L} W_{8 R}^{\mu}}{2 \sqrt{g_{L}^{2}+g_{R}^{2}}} \, , \nonumber \\
    {Z}_1^{\mu} &= \frac{\left(4 g_{L}^{2}+g_{R}^{2}\right) W_{3_L}^{\mu}-3 g_{L} g_{R} W_{3R}^{\mu}-\sqrt{3} g_{R}^{2} W_{8L}^{\mu}-\sqrt{3} g_{L} g_{R} W_{8 R}^{\mu}}{2 \sqrt{g_{L}^{2}+g_{R}^{2}} \sqrt{4 g_{L}^{2}+g_{R}^{2}}} \, , \nonumber \\
    {Z'}_2^\mu &= \frac{-g_{L} W_{8 L}^{\mu}+g_{R} W_{8 R}^\mu}{\sqrt{g_{L}^{2}+g_{R}^{2}}}  \, , \nonumber\\
    {Z'}_3^\mu &= \frac{-\left(g_{L}^{2}+g_{R}^{2}\right) W_{3 R}^{\mu}+\sqrt{3} g_{L} g_{R} W_{8L}^{\mu}+\sqrt{3} g_{L}^{2} W_{8 R}^{\mu}}{\sqrt{g_{\imath}^{2}+g_{R}^{2}} \sqrt{4 g_{L}^{2}+g_{R}^{2}}}\, , \nonumber\\
    {Z'}_4^\mu &= W_{6L}^\mu \, , \nonumber\\
    {Z'}_5^\mu &= W_{6R}^\mu \, ,
    \label{eq:ZES}
\end{align}
where $A^\mu$ is the massless eigenstate identified with the SM photon. 
The state $Z_1^\mu$ has the electroweak scale mass plus small corrections of order $(v_{ew}/V)$, and hence it is identified as the SM-like $Z^\mu$. Similarly, the mixing of ${Z'}_4^\mu$ with $( {Z'}_2^\mu, {Z'}_3^\mu, {Z'}_5^\mu)$ is also small, ${\cal O}(v_{ew}/V)$. In this limit, we identify the flavor state ${Z'}_4^\mu \equiv W_{6L}^\mu$ as the mass eigenstate ${Z}_4^\mu$. The remaining states $( {Z'}_2^\mu, {Z'}_3^\mu, {Z'}_5^\mu)$ can generally have large mixings among them. 
For simplicity, let us consider $V_2/V \ll 1$, such that  ${Z'}_5^\mu \equiv W_{6R}^\mu$ can be approximately identified with the mass eigenstate ${Z}_5^\mu$. 
The remaining two states $( {Z'}_2^\mu, {Z'}_3^\mu)$ can be easily diagonalized and their mass eigenstates are expressed as 
\begin{align}
    Z_2^\mu &= \cos\theta_0 {Z'}_2^\mu + \sin\theta_0 {Z'}_3^\mu \, ,\nonumber \\
    Z_3^\mu &= - \sin\theta_0 {Z'}_2^\mu + \cos\theta_0 {Z'}_3^\mu \, ,
    \label{eq:Z23}
\end{align}
where the mixing angle $\theta_0$ is defined as 
\begin{equation}
    \tan2\theta_0 = \frac{\sqrt{3} g_R \sqrt{4 g_L^2 + g_R^2}(2g_L^2-g_R^2 ) V_R^2}{ 2 (g_L^2 + g_R^2)^2 V^2 -3 g_R^2 (4 g_L^2 + g_R^2) V_R^2}. 
    \label{eq:neutmix}
\end{equation}
The masses of the neutral gauge bosons to the lowest order $\mathcal{O}(v_{ew}/V)$ are as follows: 
\begin{align}
    m_{Z}^{2} &= \frac{2 g_{L}^{2}\left(g_{L}^{2}+g_{R}^{2}\right)}{4 g_{L}^{2}+g_{R}^{2}}\left(S_{u u}+S_{d d}+S_{L L}-\left(S_{LV}+S_{dR }\right)^{2} / V^{2}\right)  \, ,\nonumber \\
    m_{Z_2}^2 &= \frac{1 }{3} (g_L^2+g_R^2) V^2 - \frac{1}{\sin{2\theta_0}}\frac{\sqrt{3}g_R \sqrt{4g_L^2+g_R^2} (2 g_L^2 -g_R^2)}{6 (g_L^2 + g_R^2)}V_R^2 \, , \nonumber \\
    m_{Z_3}^2 &= \frac{1 }{3} (g_L^2+g_R^2) V^2 + \frac{1}{\sin{2\theta_0}} \frac{\sqrt{3}g_R \sqrt{4g_L^2+g_R^2} (2 g_L^2 -g_R^2)}{6 (g_L^2 + g_R^2)} V_R^2\, , \nonumber \\
    m_{Z_4}^2 &= \frac{g_L^2}{2} V^2 \, , \nonumber \\
     m_{Z_5}^2 &= \frac{g_R^2}{2} V^2 \, . \label{eq:GBN}
\end{align}
The mass matrix $M_0^{'2}$ spanning the remaining basis $( W_{7L}^\mu, W_{7R}^\mu )$ is given by 
\begin{equation}
    M^{'2}_{0} = 
    \begin{pmatrix}
    \frac{g_L^2}{2} (S_{dd} + S_{LL} + V^2) & -g_L g_R (S_{dV}- S_{LR}) \\
    -g_L g_R (S_{dV}- S_{LR}) & \frac{g_R^2}{2} (S_{dd}+ S_{LL}+ V^2) \, .
    \end{pmatrix}
\end{equation}
It can be diagonalized by defining the following mass eigenstates
\begin{align}
    Z_6^\mu &= \cos\theta_0^\prime W_{7L}^\mu + \sin\theta_0^\prime W_{7R}^\mu \, ,\nonumber \\
    Z_7^\mu &= -\sin\theta_0^\prime W_{7L}^\mu + \cos\theta_0^\prime W_{7R}^\mu, 
    \label{eq:Z67}
\end{align}
where the mixing angle $\theta_0^\prime$ is defined as
\begin{equation}
    \tan2\theta_0^\prime = \frac{-4 g_L g_R (S_{dV}- S_{LR})}{(g_L^2-g_R^2) V^2}. 
    \label{eq:mix67}
\end{equation}
The masses of $( Z_7^\mu, Z_8^\mu )$ to lowest order $\mathcal{O}(v_{ew}/V)$ are given by
\begin{equation}
    M_{Z_6, Z_7}^2 = \frac{V^2}{4} \left[ g_L^2+g_R^2 \pm \frac{g_L^2-g_R^2}{\cos2\theta_0^\prime}  \right]. 
    \label{eq:GBPN}
\end{equation}
The breaking of $SU(3)_L \times SU(3)_R$ to $SU(2)_L \times U(1)_Y$ yields a relation among $\alpha_{L,R}$ and $\alpha_Y$ couplings, where $\alpha_i = g_i^2/4\pi$, and it yields \cite{Babu:2017xlu} \footnote{Also, see Ref.~\cite{Kephart:2017esj}, for a discussion of hypercharge embedding in trinification arising from D-branes.}
\begin{equation}
   \alpha_R = \frac{3}{4} \alpha_Y^{-1} - \frac{1}{4} \alpha_L^{-1}     \, .
\end{equation}
The value of $\alpha_R$ at an energy scale $\mu$ can be evaluated by solving the renormalization group equations for $\alpha_{L,Y}$, which at the one-loop level are given by 
\begin{equation}
    \alpha_i (\mu) = \frac{\alpha_i (m_t)}{1- \frac{C_i}{2\pi}\ \alpha_i (m_t)\ {\rm ln}\left(\frac{\mu}{m_t}\right)}.  
\end{equation}
Here, $\alpha_i (m_t) = 0.0102\ (0.0333)$ \cite{Buttazzo:2013uya} are the input values of $\alpha_{Y(L)}$ at $\mu = m_t = 172.44$ GeV and $C_i = 41/6\ (-19/6)$ is the beta-function coefficient of $\alpha_{Y(L)}$ from only the SM particle contributions. 
At $\mu = \cal{O}$ $(10)$ TeV, we obtain $\alpha_R/\alpha_L = 0.49$, with $g_L \simeq 0.63$. 

\begin{figure}[t!]
 \centering
\includegraphics[scale=0.5]{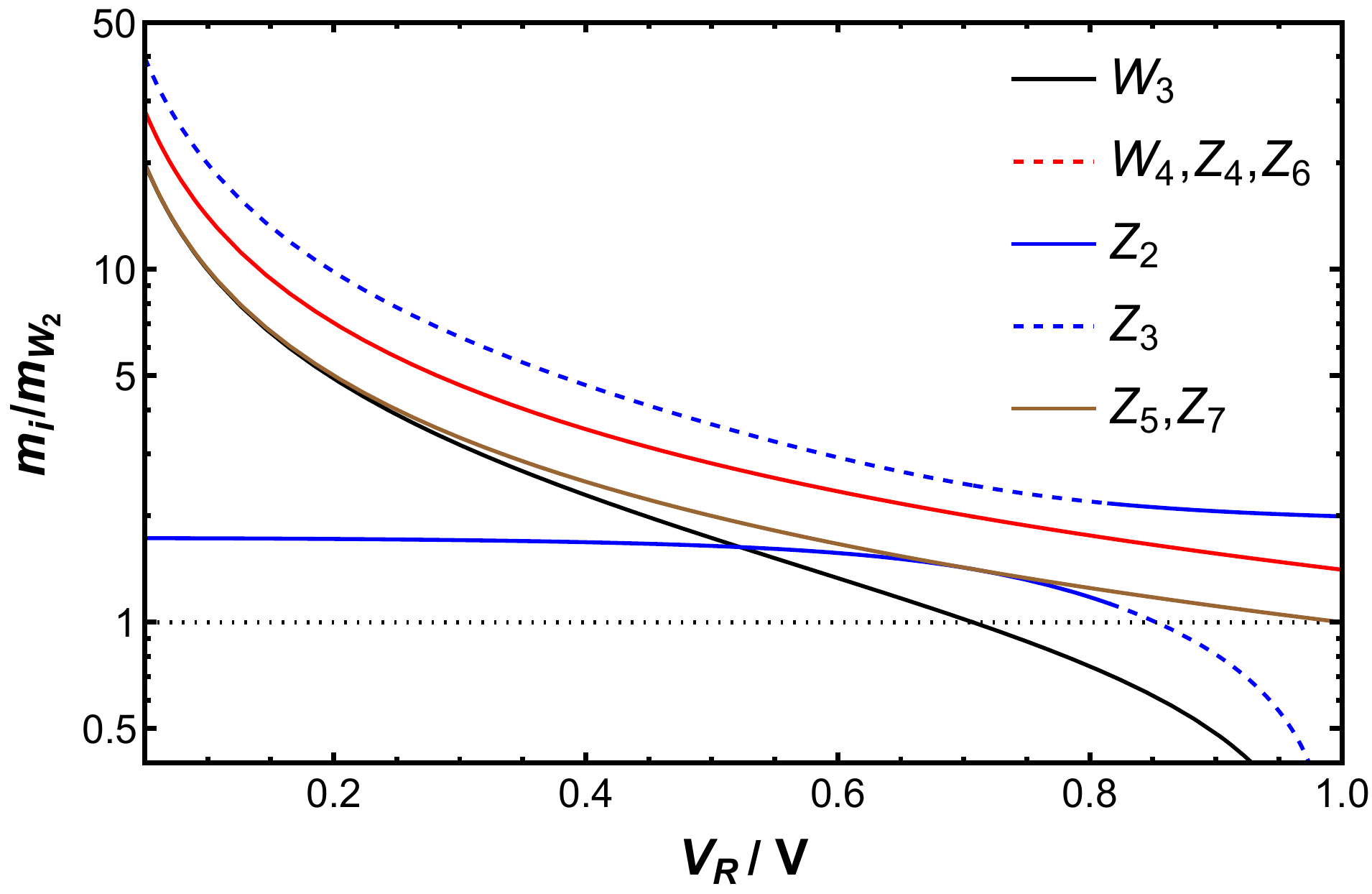}
\caption{Ratio of gauge boson masses, $m_i/m_{W_2}$ as a function of $V_R/V$ for a benchmark value $V_2/V = 10^{-3}$, $V = 10$ TeV, $g_L=0.63$, and $g_R = 0.71\ g_L$. The horizontal dotted black line corresponds to $m_i = m_{W_2}$.}
\label{fig:GBmass}
\end{figure}
Next let us now examine the relation among the gauge boson masses in the electroweak conserving limit. As observed from Eqs.~\eqref{eq:GBC}, \eqref{eq:GBN}, and \eqref{eq:GBPN}, the new gauge boson masses are all determined by $g_L$, $g_R$, $V$,  $V_2/V$ and $V_R/V$, where $g_L = 0.63$ and $g_R \simeq 0.71\ g_L \simeq 0.45$ . 
Note that the gauge boson masses for $Z_{2,3,5}$ in Eq.~\eqref{eq:GBN} are obtained  for $V_2/V\ll 1$. 
The ratios of gauge boson masses are then determined as a function of $V_R/V$ by fixing $V_2/V= 10^{-3}$. 
In Fig.~\ref{fig:GBmass}, we plot the ratio $m_i/m_{W_2}$ as a function of $V_R/V$. 
It shows that $m_{W_2}$ is the lightest gauge boson for $V_R/V \lesssim 0.50$, while $m_{W_3}$ is the lightest one for $V_R/V \gtrsim 0.7$. 
Also $m_{Z_2} < m_{Z_3}$ ($m_{Z_2} > m_{Z_3}$) for $V_R/V \lesssim 0.82$ ($V_R/V \gtrsim 0.82$) since the mixing angle $\theta_0$ between $Z_2$ and $Z_3$ in Eq.~(\ref{eq:neutmix}) flips sign above and below $V_R/V \simeq 0.82$.

\section{LHC Phenomenology and Trinification Breaking Scale}\label{sec:collider}
A TeV scale trinification symmetry breaking offers a rich phenomenology involving the new fermions and gauge bosons, which can be potentially searched for at the LHC.  
For instance, the CMS collaboration, at 95\% confidence level, has excluded down-type heavy quarks with hypercharge (-1/3) and masses below around 1500 GeV \cite{CMS:2020ttz}, as well as heavy leptons doublets with hypercharge (-1/2) with masses in the range (120 - 790) GeV \cite{CMS:2019hsm}.
There are also searches at the LHC for resonantly produced gauge boson decaying to, for example, 
(a)  top and bottom quark pair ($t{\overline b}$) \cite{ATLAS:2019fgd}, 
(b) lepton pairs, $\ell^+ \ell^-$  \cite{ATLAS:2019erb} and $e^{\pm} \nu$ \cite{ATLAS:2019lsy}, 
(c) SM $W$ or $Z$ boson and a Higgs boson \cite{ATLAS:2020qiz},  
(d) dijet with at least one isolated charged lepton \cite{ATLAS:2020zzb}, 
(e) $WW$, $ZZ$ or $WZ$ \cite{ATLAS:2020fry}.

A complete study of all these scenarios and other possibilities will be the focus of future work. 
We focus on here on the resonance production of gauge boson masses. 
The scenarios (c)$-$(e) involve the mixing between the new gauge bosons $W_i/Z_i$ and the SM $W/Z$ bosons, which is suppressed due to the hierarchy between the electroweak and the trinification breaking scale \cite{Babu:2017xlu}. 
In the following, we consider the resonance production of $W_i$ and $Z_i$ decaying to dilepton and diquark final states. 
As we discuss below, the scenarios (a) and (b) pursued by the ATLAS experiment are relevant to our case. 
The charged gauge boson in (a) correspond to the right-handed charged gauge boson of the left-right symmetric model, whereas the neutral gauge boson in (b) is the neutral gauge boson of the Sequential SM. 
In both these studies, the gauge boson couplings are fixed to be the same as the SM charged and neutral gauge boson, respectively.  
With the gauge coupling values fixed, the ATLAS collaboration at the LHC has set a lower bound on the charged and neutral gauge boson masses, $W_X \simeq 4$ TeV \cite{ATLAS:2019fgd} and $Z_X\simeq 5 $ TeV \cite{ATLAS:2019erb}, respectively.

The gauge bosons $W_{2,3}\subset ({W_R, V_R})$ and  $W_{4}\subset {V_L}$ and among them $W_R$ couples to the SM quark/lepton pairs, whereas  $V_{R,L}$ mixes the SM quarks/leptons with the heavy fermions. 
Hence, the production of $W_4$ at the LHC is highly suppressed because its production involves $d_R$ and $D_R$ mixing, which is small, $\mathcal{O} (v_{ew}/V)$. 
We find that $W_{2,3}$ only couple to right-handed fields and there is no direct interaction with $e^\pm \nu$. 
As discussed below Eq.~(\ref{eq:Z23}), for $V_2/V \ll 1$, $Z_{5}^\prime \subset W_{6R}$ decouples from  $Z_{2,3}^\prime$, and since $W_{6R}$ only has mixed-couplings with the SM quarks/leptons with the heavy fermion, $Z_{5}^\prime$ production is highly suppressed at the LHC. 
However, this is no longer true if $V_2/V \sim {\cal O} (0.1)$ for which $Z_{2,3,5}^\prime$ can mix maximally. 
For this case we evaluate the mass eigenstates by numerically diagonalizing the $Z_{2,3,5}^\prime$ mass matrix. 
For the remainder of this section, $Z_{2,3,5}$ will refer to the mass eigenstates with their masses defined to be in the order $m_{Z_1} <m_{Z_2} <m_{Z_3}$. 
Note that $Z_{4}^\prime \subset W_{6R}$ does not mix with $Z_{2,3,5}^\prime$ and    
since $W_{6L}$ only has a mixed coupling involving one SM quark/lepton and one heavy fermion, its production is highly suppressed at the LHC.

In the following we compare the resonant production cross section of  $W_{2,3}$ and $Z_{2,3,5}$ with that of $W_X$ and $Z_X$ by taking into account the difference in the gauge boson couplings to the SM fermions. 
This allows us to recast the ATLAS lower bound on gauge boson masses as  a bound on $W_2$ and $Z_{2,3,5}$ masses. 
We find that the $W_i$ and $Z_i$  couplings to the SM fermions are comparable or smaller than the SM $W$ and $Z$ boson couplings. 
The differential cross section for the resonant production of $Z_i$ or $W_i$ bosons at the LHC, $pp \to Z_i/W_i \to {\bar f_a} f_b$, where $f$'s are the SM final states, can be approximated using the narrow decay width approximation (NWA) as 
\begin{eqnarray}
  \sigma (pp \to Z_i/W_i \to {\bar f_a} f_b) = \sigma( pp \to Z_i/W_i) \times  BR(Z_i/W_i \to {\bar f_a} f_b ), 
  \label{eq:ZpLHC}
\end{eqnarray}
where $\sigma( pp \to Z_i/W_i)$  is the $Z_i/W_i$ production cross section and $BR(Z_i/W_i \to {\bar f_a} f_b )$ is the $Z_i/W_i$ branching ratio to ${\bar f_a} f_b = \ell^{+} \ell^{-} /t {\bar b}$ final states, respectively. The production and the branching ratio are expressed as  
\begin{eqnarray}
    \sigma( pp \to Z_i/W_i) \ &=& \ 2 \sum_{q, \, \bar{q}} \int^1_ {{\hat s}/s} dx \, 
  \frac{1}{x s}  
 f_q(x, Q^2)  \,  f_{\bar q} \left( \frac{{\hat s}}{x s}, Q^2
 \right) \, \hat{\sigma}(\hat{s}), 
 \nonumber \\
 BR(Z_i/W_i \to {\bar f_a} f_b)  &=& \frac{\Gamma(Z_i/W_i \to {\bar f_a} f_b)}{\Sigma_{a,b} \Gamma(Z_i/W_i \to {\bar f_a} f_b)}.  
 \label{eq:prodcs}  
\end{eqnarray}
respectively. 
Here, $f_q$ ($ f_{\bar{q}}$) is the parton distribution function (PDF) of up-type ($u$) and down-type ($d$) quarks, $\sqrt{s}=13$ TeV is the LHC Run-2 center-of-mass energy, and the production cross-section of $Z_i/W_i$ ($\hat{\sigma}$) using NWA is given by 
\begin{eqnarray}
    \hat{\sigma}(\hat{s}) \ = \ \frac{4 \pi^2}{3} \frac{\Gamma(Z_i/W_i \to q \bar{q})}{M_{i} }
   \, \delta (\hat{s}-m_{Z_2}^2), 
   \label{eq:sighat}
\end{eqnarray} 
where $M_i$ is the gauge boson mass and $\hat{s}$ is the invariant mass squared for $q {\bar q}$ pair. 
Using the production and branching ratio definitions in Eqs.(\ref{eq:prodcs}) and (\ref{eq:sighat}) and for a fixed gauge boson mass, we obtain
\begin{eqnarray}
    \frac{\sigma (pp \to Z_i/W_i \to {\bar f_a} f_b)}{\sigma (pp \to Z_X/W_X \to {\bar f_a} f_b)} \propto \frac{ \Sigma_{i} C_i \times \Gamma(Z_i/W_i \to q \bar{q})}{\Sigma_{i} C_i \times \Gamma(Z_X/W_X \to q \bar{q})} 
    \frac{BR(Z_i/W_i \to {\bar f_a} f_b)}{BR(Z_X/W_X \to {\bar f_a} f_b)},
    \label{eq:CSratio}
\end{eqnarray} 
where $C_i = 2(1)$ for $q = u(d)$ quark.

To obtain the right-hand side of Eq.~(\ref{eq:CSratio}), we have used $f_u = 2 f_d$ and $f_{\bar u} = f_{\bar d}$ to approximately account for the difference between the $u$ and $d$ quark PDFs. 
Since $g_{L,R}$ are fixed, the above ratio is determined as a function of $V$, $V_2/V$ and $V_R/V$. 
In our numerical analysis, we find that the ratio is independent of $V$. 
This can be understood in the limit $V_2/V \ll 1$; the trinification breaking scale $V$ enters the ratio through the mixing angles in Eqs.~(\ref{eq:gbC2}) and (\ref{eq:neutmix}), which are both independent of $V$. In the $Z_i$ branching ratio evaluation, we have also included the decay of $Z_i$ to electron-type heavy leptonic final states of because the heavy fermions must be lighter than the gauge bosons to explain the apparent muon $g-2$ anomaly discussed in Sec.~\ref{sec:amm}.

\begin{figure}[t!]
\includegraphics[width=0.46\textwidth]{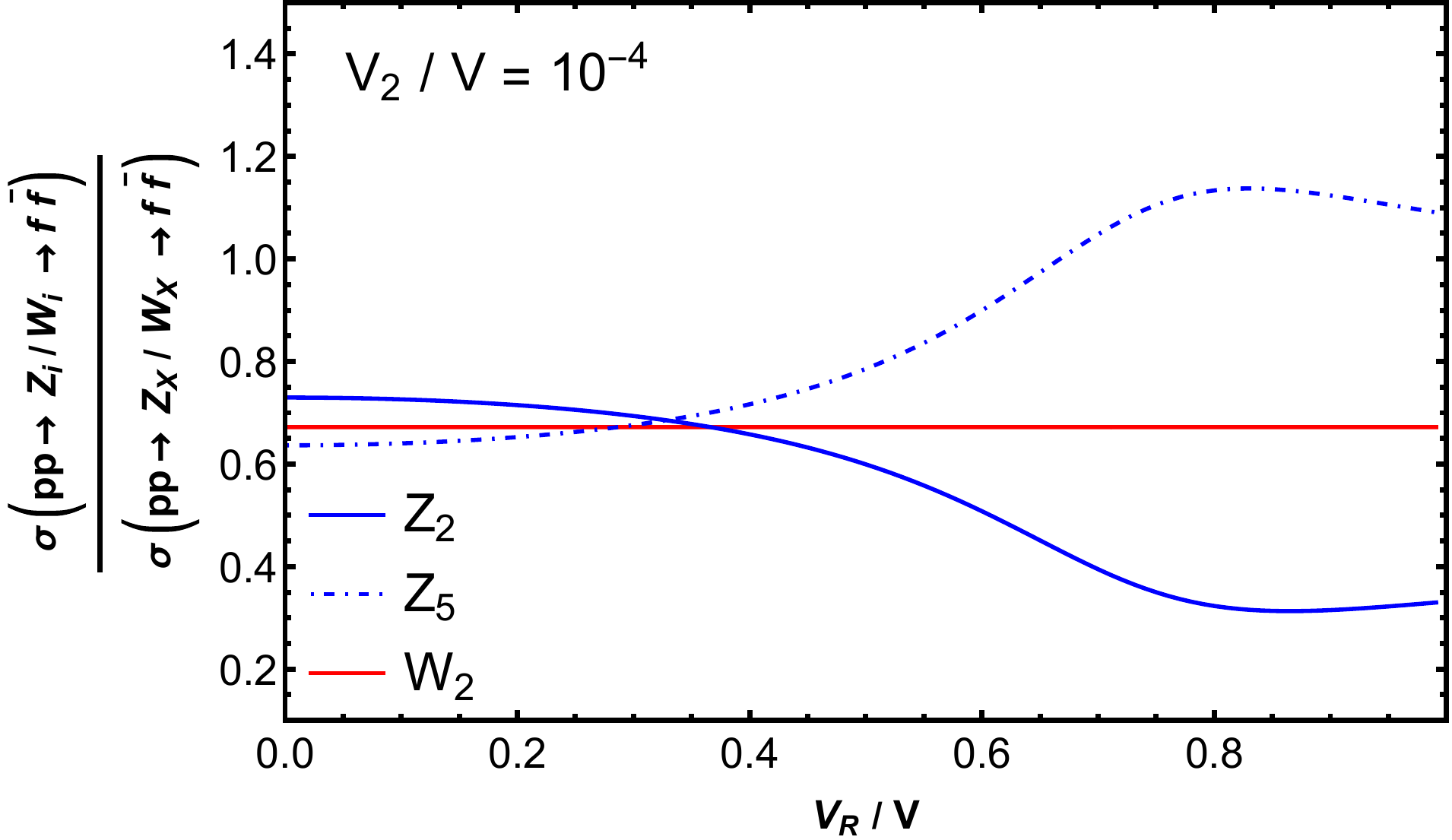}\;\;
\includegraphics[width=0.46\textwidth]{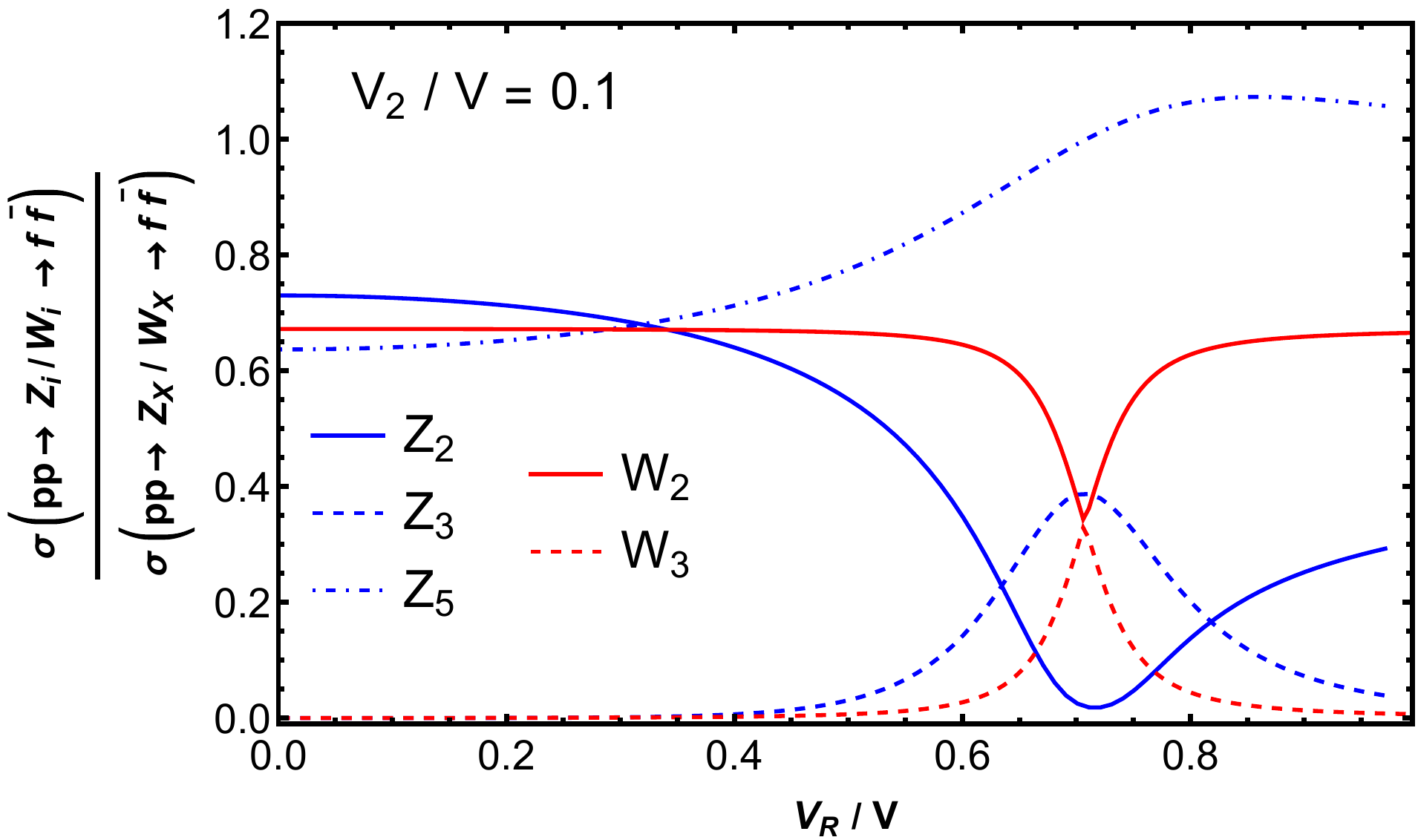}\\
\vspace{5mm}
\includegraphics[width=0.49\textwidth]{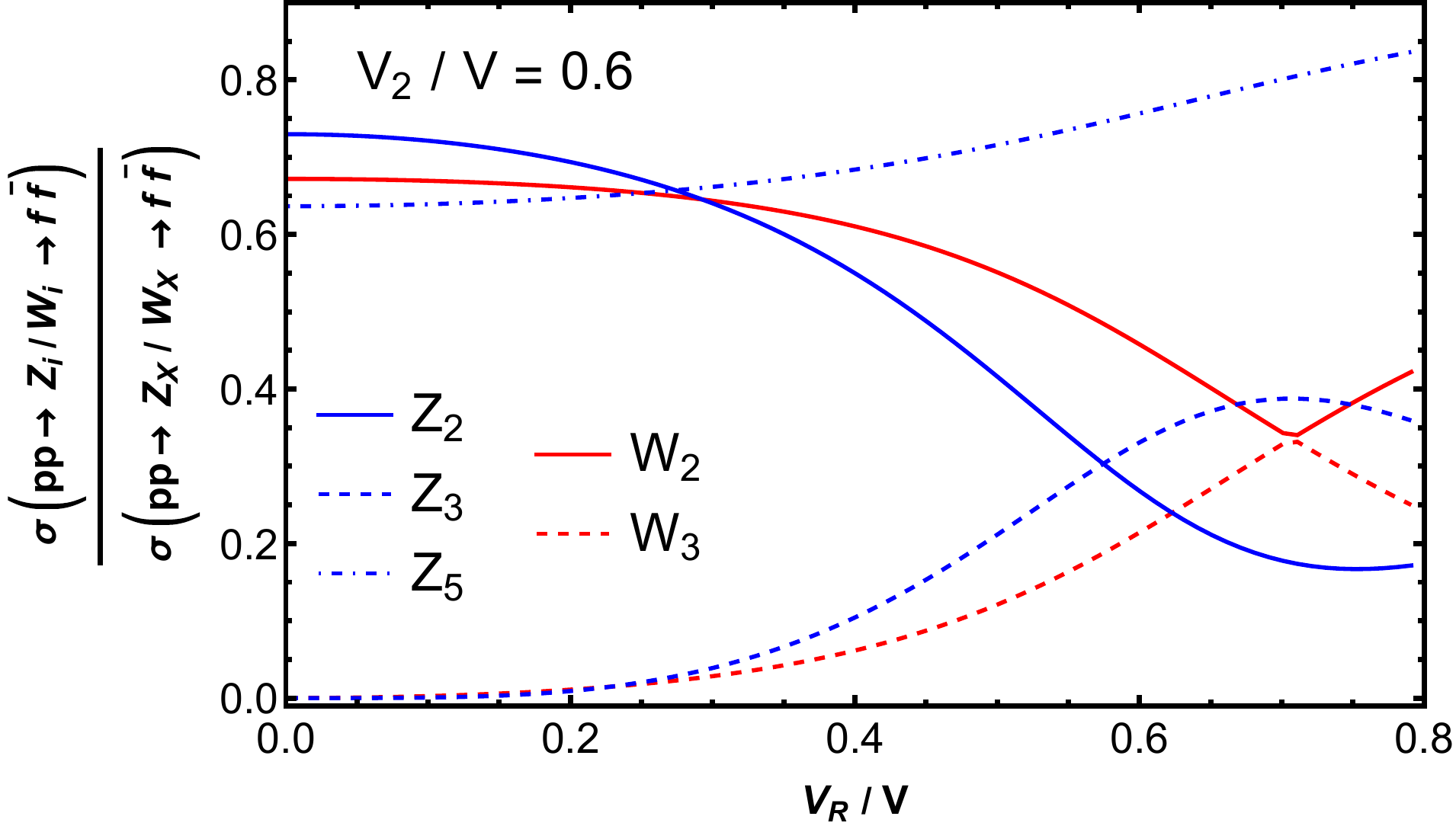}\;\;
\includegraphics[width=0.43\textwidth]{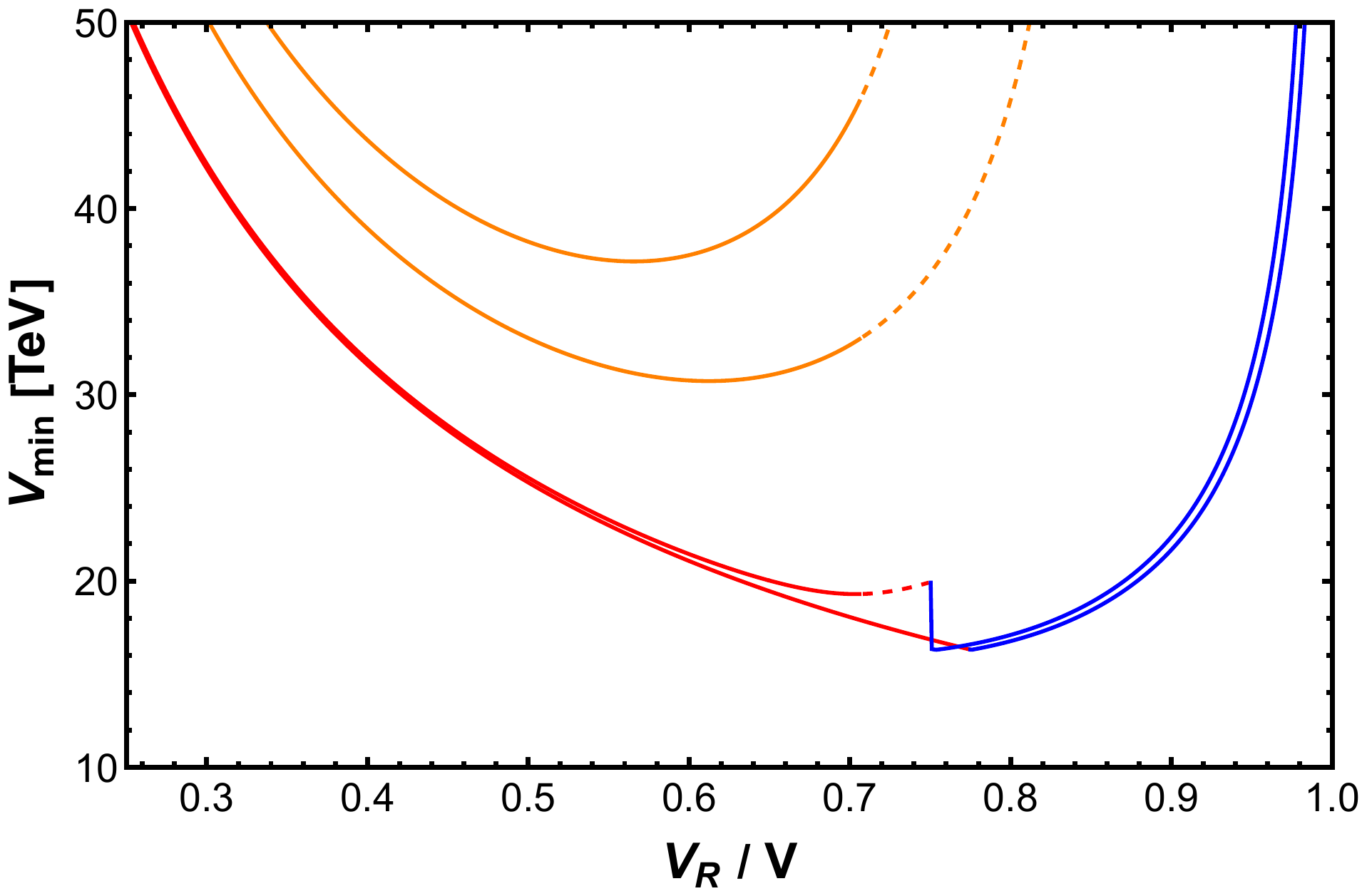}
\caption{
Top-left, top-right and bottom-left panels show the ratio of the resonance production cross sections of $W_i$ ($Z_i$) and $W_X$ ($Z_X$) bosons, which are used in LHC studies. 
We plot the ratio as  function of $V_R/V$ for fixed $g_L=0.63$, $g_R= 0.71\ g_L$ and $V_2/V$. 
For $V_2/V = 10^{-4}$ the resonance production cross section of $W_3$ and $Z_3$ is highly suppressed. 
In the bottom-right panel, we plot the lower bound on $V$ as a function of $V_R/V$. 
The curves from top to  bottom depict, $V_2/V = 0.6, 0.5,0.1, 10^{-4}$, respectively. 
The orange/red solid (dashed) lines denote the lower bounds obtained from $W_2$ ($W_3$) and the blue line represents $Z_5$, which provides the most severe bound on $V$.
The lowest value of $V$ consistent with the current LHC gauge boson resonance search bounds is for $V_2/V = 0.1$ and  $V/V_R \simeq 0.8$.}
\label{fig:collider}
\end{figure}

We show the cross section ratio for $W_i$ and $Z_i$ in the top panels and bottom-left panel of Fig.~\ref{fig:collider} as a function of $V_R/V$ for fixed $g_L=0.63$ and $g_R= 0.71\times g_L$. 
In the top-left (right) and bottom-left panel, we have fixed $V_2/V = 10^{-4}\; (0.1)$ and $0.6$. 
The $W_3$ and $Z_3$ lines are not displayed because the resonance production of $W_3$ and $Z_3$ is highly suppressed for $V_2/V = 10^{-4}$. 
For larger values of $V_2/V$, the top-right and bottom-left panel shows that $W_3$ and $Z_3$ can be produced at the LHC with cross section comparable to $W_2$ and $Z_{2,5}$, respectively. 

If the cross section ratio of $W_i$ and $Z_i$ in Fig.~\ref{fig:collider} is close to unity (for simplicity, we set this threshold value to be ${\cal O} (0.1)$ or greater), 
the lower bounds on the charged and neutral gauge boson masses by the ATLAS collaboration, $W_X \simeq 4$ TeV \cite{ATLAS:2019fgd} and $Z_X\simeq 5 $ TeV \cite{ATLAS:2019erb}, can be taken to be the lower bounds on the  $W_i$ and $Z_i$, respectively. 
We can interpret this as a bound on the trinification scale $V$,   
as shown in the bottom-right panel of  Fig.~\ref{fig:collider}. 
The curves from top to  bottom depict, $V_2/V = 0.6, 0.5,0.1, 10^{-4}$, respectively. 
The orange and red solid lines are the bounds obtained from $m_{W_3} < 4$ TeV, yields the most severe bound on $V$.  
Similarly, the dashed red and orange lines are the bounds obtained from requiring $m_{W_3} < 4$ TeV, and the blue line is the bound from requiring $m_{Z_5} < 4$ TeV. 
Comparing the minimum value of $V$ obtained for various $V_2/V$ values, we find that lowest  trinification scale $V$ consistent with the current LHC gauge boson resonance search bounds is $V \simeq  16.3$ TeV for $V_2/V = 0.1$ and  $V/V_R \simeq 0.8$. 
The HL-LHC is expected to exclude $W_X$ ($Z_X$) masses below $4.9$ ($6$) TeV \cite{CidVidal:2018eel}. 
For these values, each $V_2/V$ curve in the bottom-right panel in  Fig.~\ref{fig:collider} scales with the ratio of the expected and the current upper bounds on the gauge boson masses, namely, $\sim 5/4$ and  $6/5$ for the charged and neutral gauge boson, respectively. 

\section{Neutrino Masses}\label{sec:nuFermion}
There are five neutral leptons per generation, namely, $ (\nu, \nu^c, E^{c0}, E^0, N )$. 
The symmetric mass matrix at tree level in this basis is given by 
\begin{equation}
    \begin{pmatrix}
     0 &~~ -Y_{Ln} v_{un} &~~ 0 & -Y_{L2} V_R &~~ 0 \\
     &~~ 0 &~~ 0 & -Y_{L2} v_{L2}  &~~ 0 \\
     &~~  &~~ 0 &~~ Y_{Ln} V_n &~~ Y_{Ln} v_{un} \\
    &  &  &~~ 0 &~~ Y_{Ln} v_{dn} \\
     &  &  &  & 0
    \end{pmatrix}. 
    \label{eq:neutralMat}
\end{equation}
In the electroweak conserving limit, the neutral leptons $\nu, \nu^c,$ and $N$ have zero tree level masses per generation, implying that the extra sterile neutrinos remain light despite not being chiral under the SM. However, even in the absence of electroweak symmetry breaking, radiative corrections at the one-loop level can generate large non-zero Majorana masses for $\nu^c$ and $N$.  
The effective operator is $\Psi \Psi \Phi^\dagger \Phi^\dagger$ and, after electroweak symmetry breaking, these same operator, provide suitably light Majorana masses for the SM-like neutrinos $\nu$. 

First, let us examine masses in the electroweak conserving limit obtained by setting $v_{un}$, $v_{dn}$, and $v_{Ln}$ to zero. 
We show that four out of the five neutral leptons obtain masses without the necessity of electroweak symmetry breaking.  In this limit, the neutral lepton mass matrix given in Eq.~\eqref{eq:neutralMat} decouples into a $2N_g \times 2 N_g$ $(\nu^c, N)$ matrix and a $3 N_g \times 3 N_g$ $(\nu, E^{c0}, E^0)$ matrix where $N_g$ is the number of generations, and the mass of the heavy $SU(2)_L$ lepton doublet $(E^+, E^0)$ becomes degenerate. 
The symmetric mass matrix $M_{\nu E}$ in the basis $(\nu, E^{c0}, E^0)$ is given by  
\begin{equation}
M_{\nu E} = \begin{pmatrix}
     0 ~~& 0 &~ -Y_{L2} V_R \\
       & 0 &~ Y_{Ln} V_n  \\
       &  &~ 0 
    \end{pmatrix}. 
    \label{eq:Nmass}
\end{equation}
Ignoring generational mixing for simplicity, the above mass matrix can be digonalized by $O M_{\nu E} O^T = M_{\nu E}^{diag}$, where $M_{\nu E}^{diag}$ is the diagonal matrix in the mass basis ($\hat{\nu}, \hat{E}^{c0}, \hat{E}^0$), and the orthogonal matrix $O$ is given by 
\begin{equation}
 O = \begin{pmatrix}
  \cos\theta & \sin\theta & 0 \\
  \frac{\sin\theta}{\sqrt{2}} & \frac{-\cos\theta}{\sqrt{2}} & \frac{1}{\sqrt{2}} \\
  \frac{-\sin\theta}{\sqrt{2}} & \frac{\cos\theta}{\sqrt{2}} & \frac{1}{\sqrt{2}} \\
\end{pmatrix} \, ,  \hspace{15mm}
\tan\theta = \frac{Y_{L2}\ V_R}{Y_{Ln} V_n} \, .
\label{eq:orthN}
\end{equation}
The eigenstate $\hat{\nu}$ is massless and the pair $(\hat{E}^{c0}$, $\hat{E}^0)$ are  degenerate in mass from Eq.~\eqref{eq:dmass}, with $\hat{m}_{E^-}\equiv \hat{m}_{E}$.  
\begin{figure}[t!]
 \centering
\includegraphics[width=0.42\textwidth]{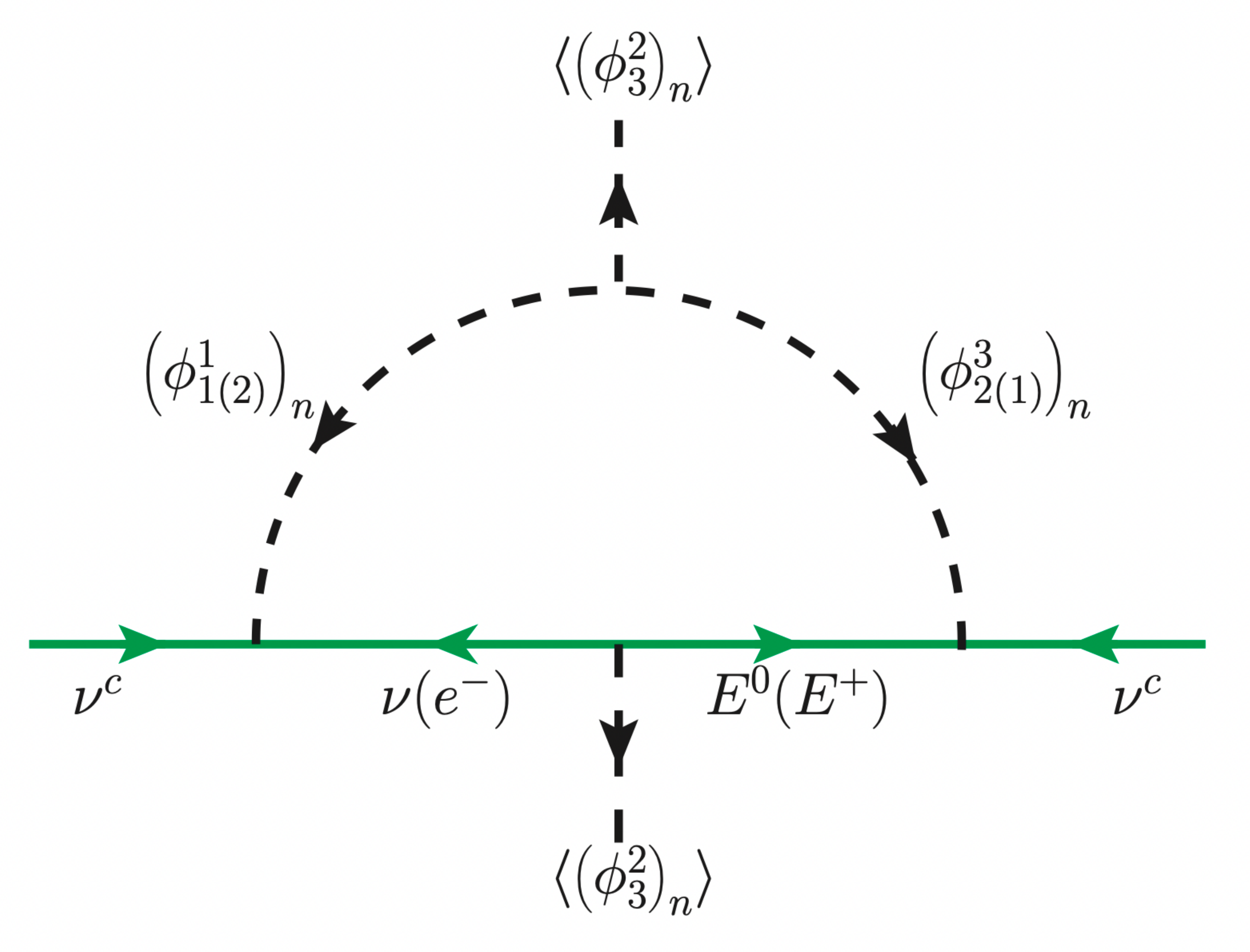}
\includegraphics[width=0.44\textwidth]{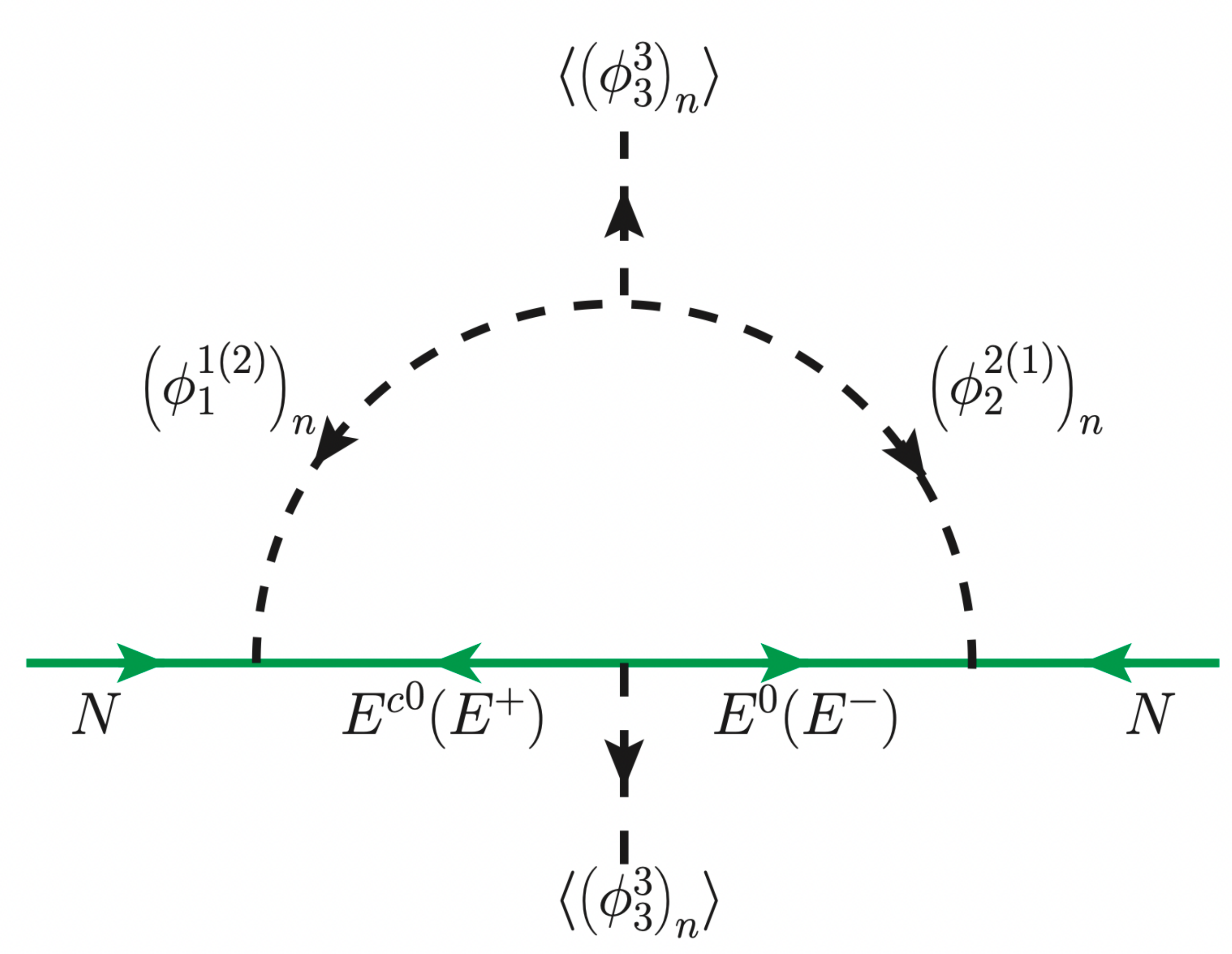}\\
(a)~~~~~~~~~~~~~~~~~~~~~~~~~~~~~~~~~~~~~~~~~~~~~~~~~~(b)\\ \vspace{3mm}
\includegraphics[width=0.42\textwidth]{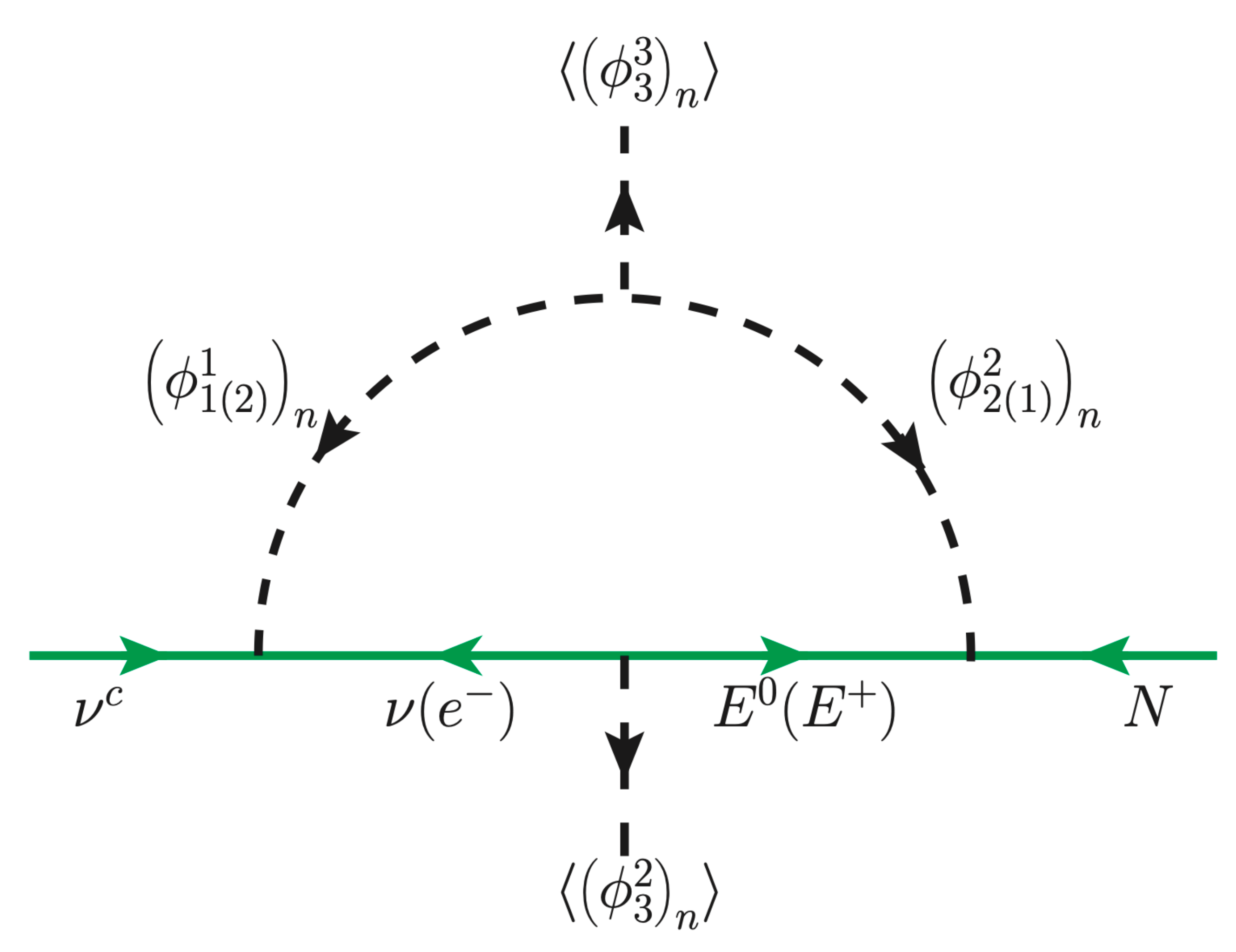}
\includegraphics[width=0.44\textwidth]{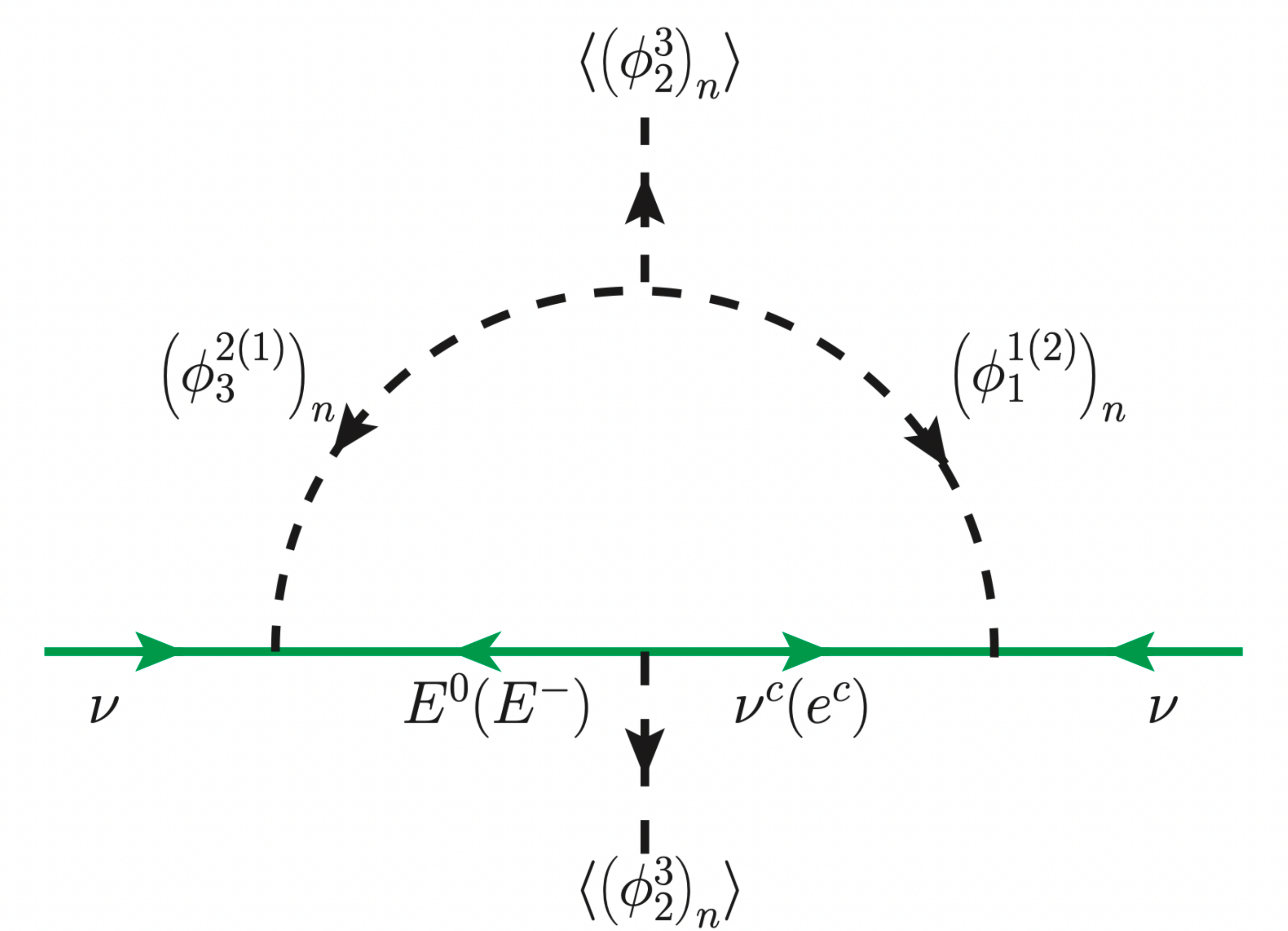}
(c)~~~~~~~~~~~~~~~~~~~~~~~~~~~~~~~~~~~~~~~~~~~~~~~~~~(d)
\caption{Radiative one-loop neutrino mass for (a) $\nu^c \nu^c$, (b) $NN$, (c) $\nu^c N$ and (d) $\nu \nu$. The generation of mass for the field $\nu$ requires electroweak symmetry breaking.  }
\label{fig:nRnR}
\end{figure}

The mass matrix $M_N^R$ spans the basis $(\nu^c, N)$ which can be expressed as
\begin{equation}
   M_N^R = \begin{pmatrix}
     m_R & m_X \\
     m_X^T & m_N
    \end{pmatrix} \, ,
    \label{eq:Nvmass}
\end{equation}
where the masses $m_R$, $m_N$, and $m_X$ are Majorana masses generated at one-loop as shown in  Figs.~\ref{fig:nRnR} (a), (b), and (c). Before evaluating these masses, let us for convenience define the mixing between $e^-$ and $E^-$ as
\begin{align}
    e_L &= \cos \theta_\ell \hat{e}_L + \sin\theta_\ell \hat{E}_L \, , \hspace{5mm}
    ~~E_L = -\sin\theta_\ell \hat{e}_L + \cos\theta_\ell \hat{E}_L  \, .
    \label{eq:eEL}
\end{align}
where the right-side of each equation is a linear combination of the mass eigenstates.

The one-loop neutrino mass represented respectively by Fig.~\ref{fig:nRnR} (a), (b), and (c) reads, 
\begin{align}
     m_R &\simeq  \{ (Y_{L1} U_{1\alpha} + Y_{L2} U_{4\alpha} )\ \hat{m}_{E}\ (Y_{L1} U_{3\alpha} + Y_{L2} U_{6\alpha} ) + T. \}\ \nonumber\\ 
     &~~~~~~~~~~~~~~~~~~~~~~~~~~~~~~~~~~~~~ \left [\frac{\sin\theta}{\sqrt{2}}\ f(\hat{m}_{E}, m_{h_\alpha^0})  + \frac{\sin 2\theta_\ell}{2} f(\hat{m}_{E}, m_{h_\alpha^+}) \right] \;,\\
     m_N &\simeq  \{ (Y_{L1} U_{1\alpha} + Y_{L2} U_{4\alpha} )\ \hat{m}_{E}\ (Y_{L1} U_{2\alpha} + Y_{L2} U_{5\alpha} ) + T. \}\  \nonumber\\ 
     &~~~~~~~~~~~~~~~~~~~~~~~~~~~~~~~~~~~~~\left[ \frac{\sin^2\theta}{2} f(\hat{m}_{E}, m_{h_\alpha^0}) + \cos\theta_\ell f(\hat{m}_{E}, m_{h_\alpha^+}) \right] \;,\\
     m_X &\simeq  \{ (Y_{L1} U_{1\alpha} + Y_{L2} U_{4\alpha} )\ \hat{m}_{E}\ (Y_{L1} U_{2\alpha} + Y_{L2} U_{5\alpha} ) + T. \}\  \nonumber\\ 
     &~~~~~~~~~~~~~~~~~~~~~~~~~~~~~~~~~~~~~\left[ \frac{\sin2\theta}{2 \sqrt{2}} f(\hat{m}_{E}, m_{h_\alpha^0}) + \sin\theta_\ell f(\hat{m}_{E}, m_{h_\alpha^+})\right] \;,
\end{align}
where $T.$ denotes transpose obtained by internal particles replacd by their charge conjugates. $\theta$ and $\theta_\ell$ are defined in Eq.~\eqref{eq:orthN} and Eq.~\eqref{eq:eEL}, and the loop integral function $f(m_a,m_b)$ is given by 
\begin{equation}
    f(m_a, m_b) =\frac{1}{4 \pi^2} \Big( \frac{m_a^2}{m_b^2-m_a^2}\ \log\Big(\frac{m_a^2}{m_b^2} \Big) + \log \Big( \frac{m_1^2}{m_b^2}\Big) \Big) \, .
    \label{eq:funcf}
\end{equation}
The mass matrix for $\{ \nu^c, N \}$ can be diagonalized by rotating to the basis parameterized as  
\begin{align}
\begin{pmatrix}
     \hat{\nu^c}  \\
     \hat{N}
    \end{pmatrix}
    =
   \begin{pmatrix}
     \cos\theta' & \sin\theta' \\
    -\sin\theta' & \cos\theta'
    \end{pmatrix}
    \begin{pmatrix}
    {\nu^c}  \\
     {N}
    \end{pmatrix} \;,
\end{align}
where the mixing angle $\theta^\prime$ is given by
\begin{equation}
    \tan2\theta' = \frac{2\ m_X}{m_R - m_N} \, .
    \label{eq:thetap}
\end{equation}
The mass eigenvalues of $\hat{\nu^c} $ and $\hat{N} $ are expressed as 
\begin{equation}
    \hat{m}_R = \frac{1}{2} \left( m_R+m_N + \frac{m_R-m_N}{\cos2\theta^\prime}  \right), \qquad\hat{m}_N = \hat{m}_N = \frac{1}{2} \left (m_R+m_N - \frac{m_R-m_N}{\cos2\theta^\prime} \right). 
    \label{eq:mRmN}
\end{equation}

It is obvious that one needs to break the electroweak symmetry in order to generate masses for the SM-like neutrino $\nu$. 
As we discussed earlier, following the electroweak breaking, a light mass for the neutrino $\nu$ is induced radiatively at  one-loop level, as shown in the Fig.~\ref{fig:nRnR} (d). 
It is given by  
\begin{equation}
    m_L \simeq \frac{1}{16 \pi^2}\ \{ (Y_{L1} + Y_{L2}) Y_{L2} (Y_{L1} + Y_{L2}) + T. \}\ \mu \; v_{L2}^2  \left(\frac{m_{\phi^+}^2 - m_{\phi^0}^2}{m_{\phi^0}^2 m_{\phi^+}^2 }\right) 
    \label{eq:t2}
\end{equation}
After taking into account the effect from the various neutral leptons states generated by the electroweak symmetry breaking as shown in Eq.~\eqref{eq:neutralMat} and ignoring the mixing between the heavy states $(\hat{E}^{c0},  \hat{E}^0)$ with $(\hat{\nu}^c, \hat{N} )$, as they are proportional to $\mathcal{O} (v_{ew}/V)$,  we obtain the following mass matrix in the $( \hat{\nu}, \hat{\nu^c}, \hat{N} )$ basis: 
\begin{equation}
   \mathcal{M} = \begin{pmatrix}
     \hat{m}_L & m_D & m_D' \\
     m_D^T & \hat{m}_R & 0 \\
     m_D'^T & 0 & \hat{m}_N. 
    \end{pmatrix}
    \label{eq:seesaw}
 \end{equation}
Here $\hat{m_R}$ and $\hat{m}_N$ are given in Eq.~\eqref{eq:mRmN}, and the matrix elements $\hat{m}_L$, $m_D$ and $m_D'$ to the leading order in $\mathcal{O} (v_{ew}/V)$ are given by 
\begin{align}
    m_D &= - (Y_{Ln} v_{un})\hat{m}_E^{-1}(Y_{Ln}V_n) \cos\theta' \nonumber\\
     m_D' &=  -m_D \tan\theta' \tan\theta \nonumber\\
     \hat{m}_L &= m_L  \hat{m}_E^{-2} (Y_{Ln}V_n)^2. 
\end{align}
To obtain the masses for the light neutrino matrix, one can make the simple assumption $m'_D \sim 0$, which reduces Eq.~\eqref{eq:seesaw} into a $6\times6$ neutrino mass matrix spanning the basis $(\hat{\nu}, \hat{\nu}^c)$. Furthermore, taking $m_D \ll \hat{m}_R$, the $3\times 3$ light neutrino mass matrix can be obtained as \
\begin{equation}
    m_{\nu}^{\rm light} = \hat{m}_L - m_D (\hat{m}_R)^{-1} m_D^T \, .
    \label{eq:lightnu}
\end{equation}
The eigenvalues of the heavier states are just equal to $\hat{m}_R$. 
Hence, in general, the light neutrino masses in this model are generated by a mixture of radiative type-I and type-II seesaw mechanisms. 
If the second (first) term in Eq.~\eqref{eq:lightnu} dominates, the neutrino mass generation is type-I (type-II). 
In the simplest scenario with $v_{L2} \sim 0$, the radiatively generated mass $m_L$ vanishes, resulting in purely the type-I scenario, whereas $m_D \sim 0$ leads to (radiatively generated) type-II dominated scenario since $\nu$ decouples from $\hat{\nu}^c$ and $\hat{N}$.

\section{Muon Anomalous Magnetic Moment}\label{sec:amm}

\begin{figure}[t!]
    \centering
    \includegraphics[scale=0.42]{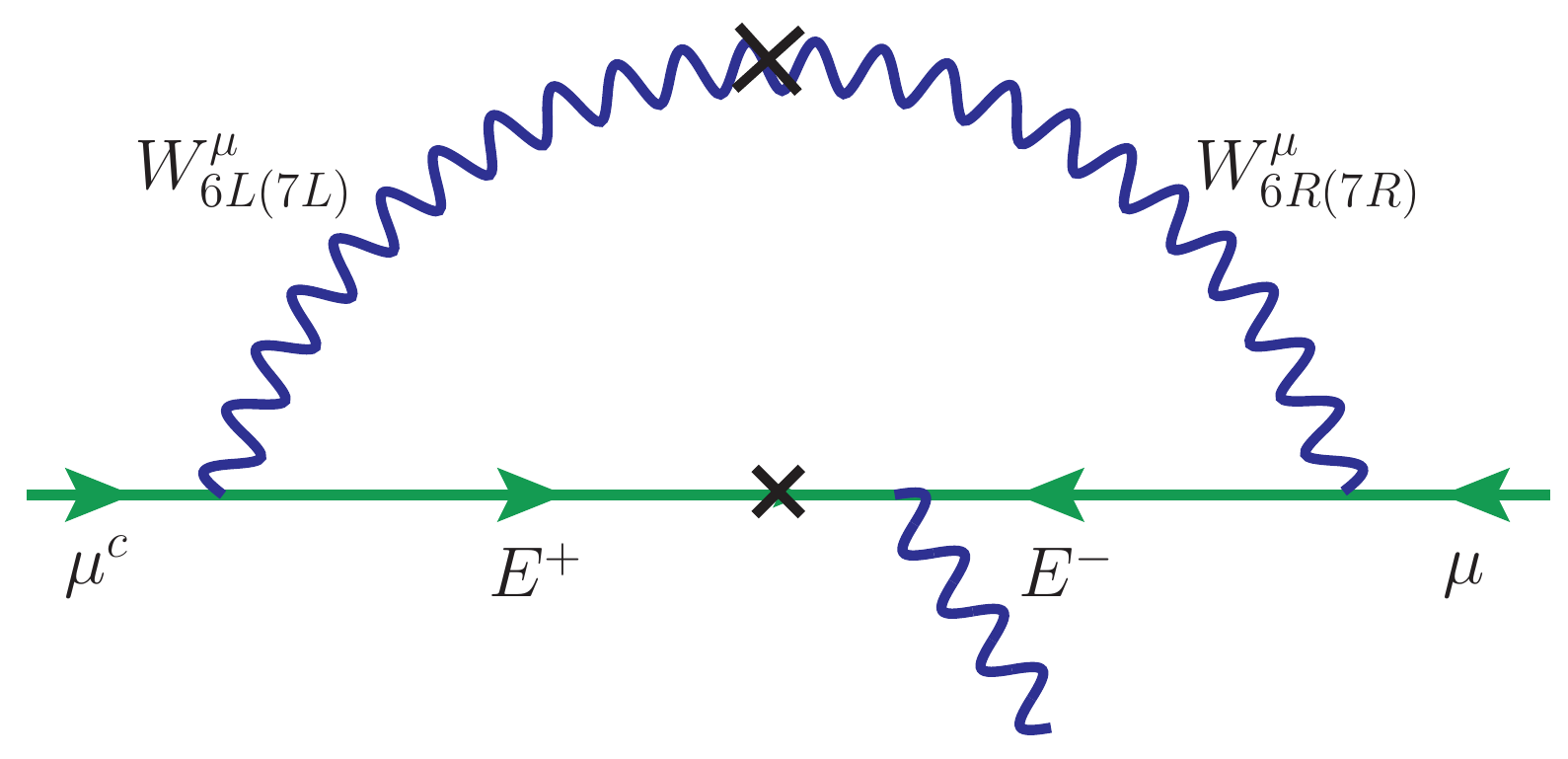} \hspace{5mm}
    \caption{A typical one-loop contribution from gauge bosons to the muon anomalous magnetic moment.}
    \label{fig:AMM}
\end{figure}

Quantum corrections arising from the interactions between the SM charged leptons, heavy leptons, and the new gauge bosons modify the electromagnetic interactions of the SM charged leptons. 
Various chirally enhanced one-loop diagrams via gauge boson mixings contribute to the anomalous magnetic moment (AMM) of the muon and electron. 
In Fig.~\ref{fig:AMM} we show a typical and most dominant Feynman diagram for $\Delta a_\mu$ involving the neutral gauge field $V^{0 \mu}_{L,R}$ exchange, which is a linear combination of left and right-handed $W_{6,7}^\mu$ fields as defined in Eq.~\eqref{eq:gaugeV}. 
Note that similar diagrams with the neutral gauge bosons replaced by charged gauge bosons $V^{\pm \, \mu}$ also exist. 
For simplicity, we suppress those diagrams by assuming adequately small mixings among the charged gauge bosons or the neutral leptons. 

\begin{figure}
    \centering
    \includegraphics[scale=0.5]{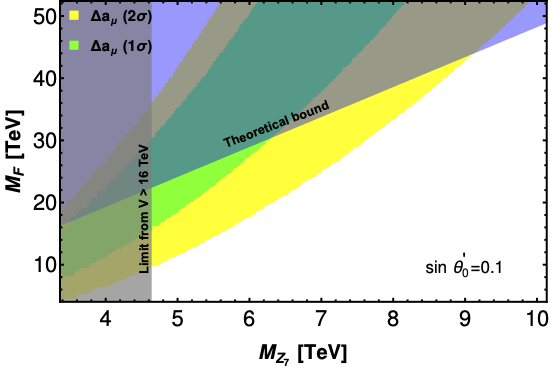}
    \caption{Allowed parameter region for muon-like heavy fermion mass $M_F$ vs. neutral gauge boson mass $M_{Z_7}$ satisfying $\Delta a_\mu$. The green (yellow) band corresponds to $\Delta a_\mu$ within $1\sigma$ ($2\sigma$). The purple band represents the exclusion region from theoretical bound ($M_F < 4.82\ M_{Z_7}$), while the gray region corresponds to the exclusion limit obtained by demanding the trinification scale $V \gtrsim$ 16 TeV.}
    \label{fig:amm}
\end{figure}

The relevant gauge interaction terms involving the neutral gauge field is given by 
\begin{equation}
    \mathcal{L} \supset m_{E^+} E^+ E^- + \frac{g_R}{2}\ \sqrt{2}\ V_R^0 \bar{E^-} \gamma_\mu e^- +  \frac{g_L}{2}\ \sqrt{2}\ V_L^{0*\mu} \bar{E^+} \gamma_\mu e^c +  h.c. \, ,
    \label{eq:lagAMM}
\end{equation}
where $V_{L,R}^0$ is defined in Eq.~\eqref{eq:gaugeV} and $m_{E^+}$ is the mass of the heavy charged lepton. Moreover, we also take the mixing between $E^-$ and the SM-like lepton $e^-$ defined in Eq.~\eqref{eq:eEL} to be small such that $E^-$ and $e^-$ are identified as the mass eigenstates. 
Using Eq.~(\ref{eq:lagAMM}), the contribution from the one-loop diagram in Fig.~\ref{fig:AMM} is given by \cite{Leveille:1977rc}
\begin{equation}
    \Delta a_\ell = - \frac{1}{4 \pi^2} \frac{m_\ell^2}{M_{X}^2} \left( (|\hat{g}_R|^2+ |\hat{g}_L|^2)\ F_1\left( \frac{M_{E_\ell^\pm}^2}{M_X^2} \right) \mp \frac{M_{E_\ell^\pm}}{m_\ell} {\rm Re} (\hat{g}_L \hat{g}_R)\ F_2\left( \frac{M_{E_\ell^\pm}^2}{M_X^2} \right) \right) \, ,
    \label{eq:AMM_1loop}
\end{equation}
where the loop functions reads 
\begin{align}
    F_1(x_\ell) &= \frac{1}{24 (x_\ell-1)^4} (8 - 38 x_\ell + 39 x_\ell^2 - 14 x_\ell^3 + 5 x_\ell^4 - 18 x_\ell^2 \ln x_\ell) \, , \\
    F_2(x_\ell) &= \frac{1}{4(x_\ell-1)^3}\ (-4+3x_\ell+x_\ell^3-6 x_\ell \ln x ) \, .
    \label{eq:AMMfun}
\end{align}
In Eq.~\eqref{eq:AMM_1loop}, $(+, -)$ corresponds to $M_X = (Z_7^\mu, Z_6^\mu)$, respectively, with the definition $(\hat{g}_R = \frac{g_R}{2} \cos\theta_0^\prime$, $\hat{g}_L = \frac{g_L}{2} \sin\theta_0^\prime )$ for eigenstate $Z_{7}^\mu$, and $(\hat{g}_R = \frac{g_R}{2} \sin\theta_0^\prime$, $\hat{g}_L = \frac{g_L}{2} \cos\theta_0^\prime)$ for eigenstate $Z_{6}^\mu$, and $\theta_0^\prime$ is the mixing angle between $(W_{7L}^\mu, W_{7R}^\mu)$ defined in Eq.~\eqref{eq:mix67}. 
The first term in Eq.~\eqref{eq:AMM_1loop} is the contribution to the AMM from the  heavy leptons without chiral enhancement, whereas the second term is the chirally-enhanced case which is proportional to the mass of the heavy muon-like lepton $E$. 
The ratio $M_{Z_6}/M_{Z_7} \approx g_R^2/g_L^2 = 0.49$ as can be seen from Eq.~\eqref{eq:mix67}. Note that the diagram with $W^\mu_{7L,7R}$ replaced by $W^\mu_{6L,6R}$ exists with the mixing angle slightly modified with $S_{LR} \to - S_{LR}$ in Eq.~\eqref{eq:mix67}. 
After considering all relevant diagrams, we find that the model can realize the desired value for muon $g-2$ as shown in Fig.~\ref{fig:amm}. 
The green (yellow) band in the figure corresponds to the allowed region in the $M_{Z_7}- M_F$ plane that can incorporate $\Delta a_\mu$ within $1\sigma$ (2$\sigma$).  The purple band represents the exclusion region from the theoretical bound obtained from mass ratio $M_F/M_{Z_7} < 4.82$ using Yukawa couplings $\leq 1.5$ (the perturbative unitarity limit on the lepton Yukawa coupling). 
The gray horizontal band is the exclusion limit obtained by demanding the trinification scale $V \gtrsim 16$ TeV. 
Thus, we find an upper limit on the gauge boson $Z_7$ mass to be 6.2 (9.0) TeV corresponding to a 30 (40) TeV heavy lepton mass to incorporate $\Delta a_\mu$ within $1\sigma$ (2$\sigma$).   

The same set of parameters also contribute to electron $g-2$ anomaly with the same sign \cite{Hanneke:2008tm, Parker:2018vye}. Hence, the model cannot simultaneously accommodate the electron and muon $g-2$ anomalies. Moreover, to satisfy the bounds from electron $g-2$ measurements, we find that the electron-type heavy lepton mass must be $\lesssim 600$ GeV, consistent with the current bounds \cite{ParticleDataGroup:2020ssz,Carpenter:2010bs}.

\section{Monopole and Fractionally Charged Leptons, Mesons and Baryons}\label{sec:monoploe}

One of the striking predictions of trinification is the presence of a topologically stable magnetic monopole that carries three quanta
$(6 \pi/e)$ of Dirac magnetic charge \cite{Kephart:2017esj, Lazarides:2021tua}. 
One expects that the monopole mass is about an order of magnitude larger than the breaking scale of the trinification gauge symmetry. Based on our discussion, this means that the monopole mass is around $160$ TeV or larger, which is orders of magnitude lighter than the superheavy GUT monopole mass. 
For a recent discussion of monopole searches at the LHC see Refs.~\cite{MoEDAL:2021mpi, ATLAS:2019wkg, HassanElSawy:2744867,Acharya:2021ckc}. 
Monopoles with a mass of the order of $10^2$ TeV can be produced from collisions of cosmic rays bombarding the Earth's atmosphere \cite{Iguro:2021xsu} and hopefully detectable in the future.

It is worth emphasizing here why the trinification monopole carries three quanta of Dirac charge. Recall that despite the presence of fractionally charged quarks, the standard GUT models based on $SU(5)$, $SO(10)$ and $E_6$ all predict the presence of a topologically stable superheavy magnetic monopole carrying a single quantum $(2 \pi/e)$ of Dirac magnetic charge. However, consistency with the Dirac quantization condition requires that this GUT monopole also carries an appropriate amount of $SU(3)_c$ color magnetic charge. There are no free quarks beyond the QCD confinement radius, and the color magnetic charge is correspondingly screened.

For completeness, it is perhaps worth pointing out here that the SM $Z$-charges carried by the quarks and leptons can also play a role in satisfying the Dirac quantization condition. This, for example, is the case for the so- called electroweak monopole identified a long time ago in $SU(2)_L \times U(1)_Y$ by Nambu \cite{Nambu:1974zg, Nambu:1977ag}. See also Refs.~\cite{Vachaspati:1992fi,Vachaspati:1992jk,Vachaspati:1994xe,Lazarides:2021los,Lazarides:2021bzg}.  
Compatibility with the Dirac quantization condition is achieved with this monopole carrying both Coulomb magnetic flux as well as $Z$- magnetic flux. The electroweak monopole does not exist as an isolated state but only in a bound   configuration together with its antimonopole (electroweak dumbbell) with mass \
$\sim$ 5-6  TeV. Such objects routinely appear in a somewhat more elaborate format in GUTs such as $SU(5)$ and $SO(10)$. See Ref.~\cite{Lazarides:2021los}
and additional references therein.

The trinification model is based on the gauge symmetry $G = SU(3)_c \times
SU(3)_L \times SU(3)_R$, which is a global direct product of the three SU(3) gauge groups. [Note that this is not the case if $G$ is embedded in $E_6$. For a recent discussion, see Ref.~\cite{Lazarides:2019xai}.]  
The observed quarks and leptons, as we have seen, reside in bi-fundamental representations of G.
However, the symmetry group G allows us to also explore exotic fermions that may reside in the fundamental representations of G, namely, $(3, 1,1)$, $(1,3,1)$, and $(1,1,3)$ together with their complex conjugates which ensures anomaly cancellation. 
With the charge operator given by Eq.~\eqref{eq:Qop},  these fermions include electrically neutral quarks (n-quarks) and fractionally charged ($\pm e/3$ and $\pm 2e/3$) f-leptons. 
In D-brane models these exotic f-leptons and n-quarks are associated with strings emitted and absorbed by the same D-brane \cite{Kephart:2017esj}. 
Although quarks are confined beyond  $\sim \Lambda_{QCD}^{-1}$, the color singlet fractionally charged fermions exist as isolated states, and therefore the trinification monopole has the required magnetic charge of $6 \pi /e$.

Interestingly, the n-quarks can combine with the SM quarks to form color singlet mesons and baryons with fractional electric charges. 
For example, the baryons $udn$ and meson $n{\bar u}$ would carry electric charges $+e/3$ and $-2e/3$, respectively. 
Thus, trinification predicts the existence of exotic fractionally charged leptons, mesons and baryons. 
The CMS collaboration has searched for color and $SU(2)_L$ singlet leptons carrying electric charge $\pm2e/3$ ($\pm e/3$) and exclude at 95\% confidence particle masses in the range [$100- 310 \;(140)$] GeV, respectively \cite{CMS:2012xi}. 
The milliQan~\cite{milliQan:2021lne} and MoEDAL \cite{Pinfold:2019zwp} experiments are expected to be more sensitive if the fractionally charged fermions have masses below $100$ GeV.


\section{Conclusion}\label{sec:conclusion}
Trinification, based on the gauge symmetry $SU(3)_c \times SU(3)_L \times SU(3)_R$, is arguably the simplest realistic extension of the SM with possible breaking at the TeV scale. 
In addition to electric charge quantization, it yields the desired value of the electroweak mixing angle with $g_R \sim 0.71g_L$, where $g_L$ and $g_R$ are the $SU(2)_L$ and $SU(2)_R$ gauge couplings, respectively. 
The masses of all the twelve new gauge bosons are uniquely determined by just three parameters associated with the trinification breaking scale, and the model also predicts the existence of new TeV scale quarks and leptons.

We have explored the phenomenology of a minimal trinification model containing just two bi-fundamental scalar fields, which  
are sufficient to accommodate the SM fermion masses, and which allows us to set a lower bound on the mass of the heaviest new d-type quark of around $15$ TeV. 
With some appropriate Yukawa coupling values, this can be interpreted as a lower bound on the trinification scale $V$. 
For more than two bi-fundamental scalar fields, the heavy quark masses are independent of SM fits, and the lower bound on $V$ is not applicable. 
A comparable bound on $V$ is obtained by considering the LHC phenomenology of the new gauge bosons. 
In particular, by examining the resonance production of the charged and neutral gauge bosons at the LHC decaying to the SM final states, we obtain a lower bound on the trinification breaking scale, $V\gtrsim 16$ TeV. 

The observed light neutrinos masses are generated via a radiative seesaw scenario, which we show is a mixture of both type-I and type-II seesaw mechanisms. 
We also find that the new gauge boson interactions involving the heavy leptons can resolve the muon $g-2$ anomaly.

Finally, there exist topologically stable magnetic monopoles carrying three quanta ($6\pi/e$) of Dirac magnetic charge and with masses about an order of magnitude larger than the symmetry breaking scale.
Monopole with $\cal{O}$ $(10^2)$ TeV scale masses could potentially be produced from collisions of cosmic rays bombarding the Earth's atmosphere. 
Consistent with the Dirac quantization condition, the trinification symmetry suggests the existence of electrically neutral quarks (n-quarks) and fractionally charged leptons (f-leptons). 
Interestingly, the n-quarks can combine with the SM quarks to form charged mesons and baryons that carry fractional charges $(\pm e/3)$ and $(\pm 2e/3)$. Together with the monopole, a search for these exotic states at high energy colliders and elsewhere remains an exciting endeavor.

\section{Acknowledgements}
D.R would like to thank N.~Okada (University of Alabama) for useful discussion on collider phenomenology. This work is supported in part by the United States Department of Energy Grants DE-SC0013880 (D.R and Q.S).

\bibliographystyle{style}
\bibliography{references}
\end{document}